\documentclass[twocolumn,aps,prc,floatfix,superscriptaddress,10pt]{revtex4-2}
\usepackage{graphicx}
\usepackage{dcolumn}
\usepackage{bm}
\usepackage[normalem]{ulem}
\usepackage{amsmath,bm}
\usepackage{amssymb}
\usepackage{longtable}
\usepackage{placeins}
\usepackage[mathlines]{lineno}
\usepackage{xcolor}
\usepackage{booktabs}
\usepackage{tabularx}
\usepackage{mathtools}
\usepackage{isotope}
\usepackage[colorlinks,linkcolor=blue,urlcolor=blue,citecolor=blue]{hyperref}
\usepackage{nccmath}
\allowdisplaybreaks[2]
\renewcommand{\sout}{\bgroup \color{red} \ULdepth=-.5ex \ULset}

\newcommand{\Ahatp}{\left( \frac{\vec{k}^{\prime}{}^2+\vec{k}{}^2}{2} \right)}

\newcommand{\Bhatp}{ \left( \vec{k}^{\prime}\cdot\vec{k} \right)}

\newcommand{\Psig}{\hat{P}_{\sigma}}

\begin{document}

\title{Anisotropic flows in Au+Au collisions at $\sqrt{s_{\rm{NN}}} = 2.4\,\text{GeV}$ with a Skyrme pseudopotential}

\author{Xin Li}
\email{Corresponding author: lixin20223@stu.htu.edu.cn}
\affiliation{College of Physics, Henan Normal University, Xinxiang 453007, China}
\affiliation{State Key Laboratory of Dark Matter Physics, Key Laboratory for Particle Astrophysics and Cosmology (MOE), and Shanghai Key Laboratory for Particle Physics and Cosmology, School of Physics and Astronomy, Shanghai Jiao Tong University, Shanghai, 200240, China}

\author{Si-Pei Wang}
\email{Corresponding author: sjtuwsp@gmail.com}
\affiliation{State Key Laboratory of Dark Matter Physics, Key Laboratory for Particle Astrophysics and Cosmology (MOE), and Shanghai Key Laboratory for Particle Physics and Cosmology, School of Physics and Astronomy, Shanghai Jiao Tong University, Shanghai, 200240, China}
\affiliation{College of Physics, Henan Normal University, Xinxiang 453007, China}

\author{Rui Wang}
\email{Corresponding author: rui.wang@lns.infn.it}
\affiliation{Istituto Nazionale di Fisica Nucleare (INFN), Laboratori Nazionali del Sud, I-$95123$ Catania, Italy
}%
\author{Zhen Zhang}
\email{Corresponding author: zhangzh275@mail.sysu.edu.cn}
\affiliation{Sino-French Institute of Nuclear Engineering and Technology, Sun Yat-sen University, Zhuhai 519082, China}

\author{Jie Pu}
\email{Corresponding author: wlpujie@126.com}
\affiliation{College of Physics, Henan Normal University, Xinxiang 453007, China}
\affiliation{Institute of Nuclear Science and Technology, Henan Academy of Science, Zhengzhou, 450046, China}
\affiliation{Shanghai Research Center for Theoretical Nuclear Physics, NSFC and Fudan University, Shanghai, 200438, China}

\author{Chun-Wang Ma}
\email{Corresponding author: machunwang$@$126.com}
\affiliation{College of Physics, Henan Normal University, Xinxiang 453007, China}
\affiliation{Institute of Nuclear Science and Technology, Henan Academy of Science, Zhengzhou, 450046, China}
\affiliation{Shanghai Research Center for Theoretical Nuclear Physics, NSFC and Fudan University, Shanghai, 200438, China}

\author{Lie-Wen Chen}
\email{Corresponding author: lwchen$@$sjtu.edu.cn}
\affiliation{State Key Laboratory of Dark Matter Physics, Key Laboratory for Particle Astrophysics and Cosmology (MOE), and Shanghai Key Laboratory for Particle Physics and Cosmology, School of Physics and Astronomy, Shanghai Jiao Tong University, Shanghai, 200240, China}

\date{\today}

\begin{abstract}
Anisotropic flows in heavy-ion collisions provide a basic experimental observable to understand the nuclear
collision dynamics and to constrain dense nuclear matter equation of state~(EOS).
Within the framework of the lattice Boltzmann-Uehling-Uhlenbeck transport model,
we present a systematic study of proton anisotropic flow observables measured by the HADES collaboration, by utilizing the recently developed nuclear effective interaction based on the density-, momentum- and isospin-dependent N$5$LO Skyrme pseudopotential.
In particular, we investigate the impacts of the momentum dependence of nucleon mean-field potentials, the stiffness of symmetric nuclear matter~(SNM) EOS, the high-density behaviors of the symmetry energy and the in-medium modification of nucleon-nucleon elastic cross sections on proton directed~($v_{1}$), elliptic~($v_{2}$), triangular~($v_{3}$), and quadrangular~($v_{4}$) flows in Au+Au collisions at $\sqrt{s_{\rm{NN}}} = 2.4\,\text{GeV}$.
Our results show that the proton anisotropic flows are strongly sensitive to the momentum dependence of nucleon mean-field potential as well as the incompressibility coefficient $K_0$ of SNM. In addition, the transverse momentum dependence of the proton $v_2$ exhibits a modest sensitivity to the higher-order skewness coefficient $J_0$ and kurtosis coefficient $I_0$ of SNM as well as the momentum dependence of the symmetry potential, while the transverse momentum dependence of the proton $v_1$ is shown to modestly depend on the in-medium modification of nucleon-nucleon elastic cross sections. Moreover, the high-density symmetry energy seems to have limited effects on the proton anisotropic flows.
These findings highlight the necessity of considering the
momentum dependence of nucleon mean-field potentials including the symmetry potential,
the higher-order characteristic parameters of SNM EOS beyond $K_0$, and the in-medium modification of nucleon-nucleon elastic cross sections, in future Bayesian transport model analyses on proton anisotropic flows in heavy-ion collisions at HADES energies,
thereby to extract information on nuclear matter EOS as well as the associated underlying nuclear effective interactions.
\end{abstract}

\maketitle


\section{Introduction}
\label{introction}
The exploration of nuclear matter equation of state (EOS) represents a fundamental objective in nuclear physics,
as it provides crucial insights into the properties of nuclear matter under extreme
conditions, which is valuable for both finite nuclei, nuclear reactions, and nuclear astrophysics~\cite{Li:1997px,Lattimer:2000kb,Danielewicz:2002pu,Lattimer:2004pg,Baran:2004ih,Steiner:2004fi,
Li:2008gp,Oertel:2016bki,Sorensen:2023zkk}. Since the mid-1980s, the EOS of symmetric nuclear matter has been relatively well-constrained near nuclear saturation density ($\rho_0$) from nuclear giant monopole resonance
experiments ~\cite{Blaizot:1980tw,Youngblood:1999zza,Li:2007bp,Garg:2018uam,Li:2022suc}.
Its behavior at suprasaturation densities has also been explored through heavy-ion collisions~\cite{Aichelin:1985rbt,Fuchs:2000kp,Li:1997px,Danielewicz:2002pu,Baran:2004ih,Li:2008gp,Hartnack:2005tr,Fuchs:2005zg,LeFevre:2015paj,Sorensen:2023zkk,Cozma:2024cwc}.
However, the isospin-dependent component of nuclear matter
EOS, governed by the symmetry energy \( E_{\rm sym}(\rho) \), remains highly uncertain, particularly at suprasaturation densities~\cite{FiorellaBurgio:2018dga,
Zhang:2018bwq,Zhou:2019omw,Zhou:2019sci,Yue:2021yfx,Li:2021thg,Li:2025xio}.
Recent multi-messenger astronomical observations have shed new light on constraining the properties of nuclear matter, particularly the EOS on dense and neutron-rich systems~\cite{Zhou:2019omw,Zhou:2019sci,Yue:2021yfx,Li:2021thg,Cao:2023rgh,Tsang:2023vhh,Qiu:2023kfu,Yue:2024srj,Koehn:2024set,Ye:2024meg,Qiu:2025kyv,Steinheimer:2025hsr,Cai:2025nxn,Li:2025xio}, and also have highlighted the crucial interplay among nuclear theory, laboratory experiments, and astrophysical observations.

In addition to the nuclear matter EOS, another quantity that plays a crucial role in nuclear physics, especially for dynamic processes such as heavy-ion collisions and neutron star mergers, is the nucleon single-particle potential (mean-field potential), which encodes the basic properties of the underlying nuclear effective interactions.
Experimental determinations of this quantity can be achieved through optical model analyses of nucleon scattering data.
Currently, the behaviors of nucleon single-particle potentials in nuclear medium remains largely uncertain, even at the nuclear saturation density.
Different non-relativistic reductions from relativistic optical potential can lead to different single-particle potentials, for example, the Hama potential~\cite{Hama:1990vr,Cooper:1993nx} and the one obtained by Feldmeier and Lindner~\cite{Feldmeier:1991ey}.
The situation becomes even more uncertain when it comes to the isospin-dependent part of the nucleon single-particle potential, i.e., the symmetry potential, which is closely related to the neutron–proton effective mass splitting. Optical model analyses on the symmetry potential have so far been limited to low nucleon momenta, and consensus---even qualitatively---on the neutron-proton effective mass splitting at nuclear saturation density is still lacking~\cite{Xu:2010fh,Li:2014qta,Li:2018lpy}.

Heavy-ion collisions provide a unique platform for exploring the properties of nuclear matter under extreme conditions of density, temperature, and isospin asymmetry~\cite{Sorensen:2023zkk}, as well as the nucleon single-particle behaviors~\cite{Li:2018lpy}.
Among experimental probes of heavy-ion collisions, collective flows stand out as established and reliable probes for extracting the properties of dense nuclear matter, due to their sensitivity to the EOS
~\cite{Stoecker:1986ci,Gale:1987zz,Li:1996ix,Pak:1997zza,Chen:1998wct,Danielewicz:1998vz,Danielewicz:1999zn,Scalone:1999mwx,Li:2000bj,Danielewicz:2002pu,Giordano:2010pv,Nara:2020ztb,Nara:2021fuu,Cozma:2024cwc,Chen:2024aom,Liu:2025pzr}.
Recent high-precision measurements of anisotropic flows in Au+Au collisions at $\sqrt{s_{\rm NN}}$ $=$ $2.4~\rm GeV$ (corresponding to an incident beam energy of $E_{\rm beam}$ $=$ $1.23A~\rm GeV$) by the HADES Collaboration~\cite{HADES:2020lob,HADES:2022osk} have opened new possibilities for probing the nuclear matter EOS at suprasaturation densities. However, establishing connections between these experimental observables and certain properties of nuclear matter presents significant challenges.
Anisotropic flows exhibit sensitivities not only to the nuclear matter EOS, but also to other properties such as the nucleon single-particle potential, the in-medium nucleon-nucleon elastic cross sections, and the detailed collision dynamics~\cite{Li:2005jy,Isse:2005nk,Nara:2020ztb,Nara:2021fuu,Li:2022wvu,Parfenov:2022brq,Li:2023ydi,Du:2023ype,Liu:2023rlm,Kireyeu:2024hjo,Mohs:2024gyc,Steinheimer:2024eha,Steinheimer:2025hsr,Li:2025iqq,Liu:2025pzr,Bratkovskaya:2025oys}.
This multi-factor dependence introduces substantial uncertainties in microscopic dynamical approaches, highlighting the need for reliable theoretical frameworks of heavy-ion collisions and systematic uncertainty analyses.

A detailed understanding of collision dynamics is essential for describing heavy-ion collisions and extracting relevant information from experimental data. To this end, the microscopic transport models such as the Boltzmann-Uehling-Uhlenbeck (BUU) equation ~\cite{Bertsch:1988ik} and the quantum molecular dynamics (QMD) model~\cite{Aichelin:1991xy} have been developed and widely applied. However, due to the presence of model dependencies and uncertainties in transport model simulations, the Transport Model Evaluation Project~\cite{TMEP:2016tup,TMEP:2017mex,TMEP:2021ljz,TMEP:2022xjg,TMEP:2019yci,TMEP:2023ifw} has also been pursued to further quantify uncertainties and therefore enhance the predictive power of transport models (see also Ref.~\cite{Kolomeitsev:2004np}). In fact, one of the basic inputs in the microscopic transport models is the non-equilibrium nucleon single-particle potentials (as a functional of the one-body nucleon occupation or Wigner function), which can be derived from a given nuclear effective interaction based on the nuclear energy density functional (EDF).
Since the nuclear matter EOS can be obtained based on the same nuclear effective interaction, the exact information of the nuclear matter EOS is therefore encoded into transport models.
By comparing model calculations with experimental data, information on the nuclear matter EOS and the nucleon single-particle behaviors in nuclear matter can then be obtained.
In order to delineate the sensitivity of certain observables in heavy-ion collisions to particular characteristic quantities of the nuclear matter EOS or nucleon single-particle potential, a flexible nuclear effective interaction, which can disentangle the correlations between different characteristic quantities, is required.
One of such interactions is the Skyrme pseudopotential, which is a generalization of the conventional Skyrme effective interaction by incorporating higher-order momentum derivative terms~\cite{Carlsson:2008gm,Raimondi:2011pz}.
It has been extended to include momentum derivative terms up to the tenth order, i.e., N$5$LO Skyrme pseudopotential, to reproduce the empirical nucleon optical potential up to energy of $2~\rm GeV$~\cite{Wang:2024xzq}.
Combining with the density-dependent (DD) terms extended through a Fermi-momentum expansion approach~\cite{Patra:2022yqc}, the N$5$LO Skyrme pseudopotential can provide high flexibility on both momentum dependence of nucleon single-particle potential in nuclear matter and the density dependence of nuclear matter EOS.

In this work, we aim to investigate the effects of the momentum dependence of nucleon single-particle potential, the stiffness of the isospin symmetric part of the nuclear matter EOS, the high-density behaviors of the symmetry energy, and the in-medium modification of nucleon-nucleon elastic cross sections on proton anisotropic flows in Au+Au collisions at HADES energies, within the lattice BUU~(LBUU) transport model using the N$5$LO Skyrme pseudopotential~\cite{Wang:2024xzq}. We regard this systematic investigation as a foundation for further extracting the information on the nuclear matter EOS and nucleon single-particle behaviors employing comprehensive uncertainty analysis methods such as the Bayesian inference.
In Sec.~\ref{Methodology}, we introduce the Skyrme pseudopotential, the LBUU transport model, and describe the interactions employed, along with the basic concept of anisotropic flow.
In Sec.~\ref{discussion}, we present the anisotropic flow coefficients with different nuclear matter EOSs and nucleon single-particle potentials, and discuss their respective impacts.
Finally, a summary and brief outlook are provided in Sec.~\ref{summary}.

\section{Theoretical framework}
\label{Methodology}

\subsection{Skyrme pseudopotential and nuclear matter equation of state}
\label{Skyrme}

Effective interactions involving quasilocal operators depending on
spatial derivatives are conventionally called as pseudopotential. A general method for constructing the Skyrme-like quasi-local EDF has been proposed in Refs.~\cite{Carlsson:2008gm,Raimondi:2011pz}, where higher-order derivative terms are incorporated in the conventional Skyrme effective interaction~\cite{Chabanat:1997qh,Chabanat:1997un}.

In previous works~\cite{Carlsson:2008gm,Raimondi:2011pz}, the N3LO Skyrme pseudopotential has been constructed by introducing the additional fourth- and sixth-order derivative terms into the conventional Skyrme interaction.
The central term of the N3LO Skyrme pseudopotential is written as
\begin{widetext}
\begin{equation}
	\label{eq:VN3LO}
	\small
	\begin{aligned}
		V_{\mathrm{N} 3 \mathrm{LO}}^C  =
		& t_0\left(1+x_0 \hat{P}_\sigma\right)+t_1^{[2]}\left(1+x_1^{[2]} \hat{P}_\sigma\right) \frac{1}{2}\left(\hat{\vec{k}}^{\prime 2}+\hat{\vec{k}}^2\right)+t_2^{[2]}\left(1+x_2^{[2]} \hat{P}_\sigma\right) \hat{\vec{k}}^{\prime} \cdot \hat{\vec{k}}+t_1^{[4]}\left(1+x_1^{[4]} \hat{P}_\sigma\right)\left[\frac{1}{4}\left(\hat{\vec{k}}^{\prime 2}+\hat{\vec{k}}^2\right)^2+\left(\hat{\vec{k}}^{\prime} \cdot \hat{\vec{k}}\right)^2\right] \\
		& +t_2^{[4]}\left(1+x_2^{[4]} \hat{P}_\sigma\right)\left(\hat{\vec{k}}^{\prime} \cdot \hat{\vec{k}} \right)\left(\hat{\vec{k}}^{\prime 2}+\hat{\vec{k}}^2\right)+t_1^{[6]}\left(1+x_1^{[6]} \hat{P}_\sigma\right)\left(\hat{\vec{k}}^{\prime 2}+\hat{\vec{k}}^2\right)\left[\frac{1}{2}\left(\hat{\vec{k}}^{\prime 2}+\hat{\vec{k}}^2\right)^2+6\left(\hat{\vec{k}}^{\prime} \cdot \hat{\vec{k}}\right)^2\right] \\
		& +t_2^{[6]}\left(1+x_2^{[6]} \hat{P}_\sigma\right)\left(\hat{\vec{k}}^{\prime} \cdot \hat{\vec{k}}\right)\left[3\left(\hat{\vec{k}}^{\prime 2}+\hat{\vec{k}}^2\right)^2+4\left(\hat{\vec{k}}^{\prime} \cdot \hat{\vec{k}}\right)^2\right] ,
	\end{aligned}
\end{equation}
where $\hat{P}_\sigma$ is the spin-exchange operator;
$ \hat{\vec{k}}=-i\left( \hat{\vec{\nabla}}_1-\hat{\vec{\nabla}}_2 \right)/2 $ is the relative momentum operator and $\hat{\vec{k}}^{\prime} $ is the conjugate operator of $ \hat{\vec{k}}$ acting on the left.
The corresponding Hamiltonian density and single-nucleon potential under general nonequilibrium conditions has been derived in Ref.~\cite{Wang:2018yce,Wang:2023gta} within Hartree-Fock approximation. It has been shown that the N$3$LO Skyrme pseudopotential~\cite{Wang:2018yce,Wang:2023gta}  can reproduce the empirical nucleon optical potential up to $1$~GeV~\cite{Hama:1990vr,Cooper:1993nx}.
In Ref.~\cite{Wang:2024xzq}, the Skyrme pseudopotential is further extended up to the N5LO by including the additional eighth- and tenth-order derivative terms:
\begin{equation}
	\label{eq:v8_new}
	\small
	\begin{aligned}
		v^{[8]} = & \, t_{1}^{[8]}\left( 1 + x_{1}^{[8]} \Psig \right) \left[ \Ahatp^4 + 6 \Ahatp^2 \Bhatp^2 + \Bhatp^4 \right]  \\
		& + t_{2}^{[8]}\left( 1 + x_{2}^{[8]} \Psig \right) \left[ 4 \Ahatp^3 \Bhatp  + 4 \Ahatp \Bhatp^3 \right] ,
	\end{aligned}
\end{equation}
and
\begin{equation}
	\label{eq:v10_new}
	\small
	\begin{aligned}
		v^{[10]} = & \, t_{1}^{[10]}\left( 1 + x_{1}^{[10]} \Psig \right) \left[ \Ahatp^5 + 10 \Ahatp^3 \Bhatp^2 + 5 \Ahatp \Bhatp^4      \right] \\
		& + t_{2}^{[10]}\left( 1 + x_{2}^{[10]} \Psig \right) \left[ 5\Ahatp^4 \Bhatp + 10 \Ahatp^2 \Bhatp^3 + \Bhatp^5  \right].
	\end{aligned}
\end{equation}
This extension leads to the inclusion of higher-order momentum dependencies, i.e., $p^8$ and $p^{10}$, into the single-nucleon potential, enabling it not only reproduce the empirical nucleon optical potential up to $1$~GeV but also maintain saturation up to $2$~GeV~\cite{Wang:2024xzq}.
Therefore, the N5LO Skyrme pseudopotential can be safely applied in the simulations of Au$+$Au collision at HADES energies~\cite{HADES:2020lob,HADES:2022osk}.
Thus the central term of the N5LO Skyrme pseudopotential used in this work can be expressed as
\begin{equation}
	V_{\mathrm{N5LO}}^{C}=V_{\mathrm{N3LO}}^{C}+v^{[8]}+v^{[10]}.
\end{equation}
Based on the Fermi momentum expansion proposed in Ref.~\cite{Wang:2023zcj,Wang:2024xzq}, the DD term can be written as
\begin{equation}
	\label{eq:V_DD}
	V^{\mathrm{DD}}_{N} = \sum_{n=1}^{N} \frac{1}{6} t_{3}^{[2n-1]} \left( 1+ x_{3}^{[2n-1]} \Psig \right) \rho^{\frac{2n-1}{3}} ( \vec{R} ),
\end{equation}
where $\vec{R}=\left( \vec{r}_1 + \vec{r}_2 \right)/2$.
For brevity, the factor $\hat{\delta}\left(\vec{r_1}-\vec{r_1} \right)$ is omitted from Eqs.~(\ref{eq:VN3LO}), (\ref{eq:v8_new}), (\ref{eq:v10_new}) and (\ref{eq:V_DD}).
It is evident that increasing the number of DD term enhances the flexibility of the density dependence of both SNM EOS and symmetry energy.
Here, We choose $N=4$ to provide sufficient flexibility without introducing excessive parameters, and the Skyrme pseudopotential N$5$LODD$4$ used in this work are then written as
\begin{equation}
	v^{\mathrm{N5LODD4}}_{sk}= V_{\mathrm{N5LO}}^C + V_{4}^{\mathrm{DD}}.
\end{equation}
In the above, the $t_0$, $x_0$; $t_{i}^{[n]}$, $x_{i}^{[n]}$ ($n=2,4,6,8,10$ and $i=1,2$); $t_{3}^{[n]}$, $x_{3}^{[n]}$ ($n=1,3,5,7$) are $30$ Skyrme parameters.

\end{widetext}

Most generally, for a given two-body nuclear effective interaction $v_{12}$, the Hamiltonian density ${\cal H}(\mathbf{r})$ can be expressed as a functional of the nucleon one-body phase-space distribution functions $f_\tau(\mathbf{r},\mathbf{p})$, with $\tau=1$ (or n) for neutrons and $-1$ (or p) for protons.
The nucleon single-particle potential $U_\tau(\mathbf{r},\mathbf{p})$ is obtained by taking variations of the potential part of ${\cal H}(\mathbf{r})$ with respect to $f_\tau(\mathbf{r},\mathbf{p})$.
By evaluating $f_\tau(\mathbf{r},\mathbf{p})$ as the zero-temperature Fermi distribution, we obtain the nucleon single-particle potential $U_\tau(\rho,\delta,p)$, which reduces to an analytical function of density ($\rho$), isospin asymmetry ($\delta$) and the magnitude of nucleon momentum $|\vec{p}|$ in cold nuclear matter.
Within the framework of N$5$LODD$4$ Skyrme pseudopotential, $U_\tau(\rho,\delta,p)$ in symmetric nuclear matter at nucleon saturation density, $U_0(\rho_0,p)$, can be expressed in a polynomial form~\cite{Wang:2024xzq},
\begin{equation}
	\label{eq:U0_N5LO}
	\begin{aligned}
		U_{0}(\rho_{0},p) =
		~&a_{0}
		+ a_{2} \left(\frac{p}{\hbar}\right)^{2}
		+ a_{4} \left(\frac{p}{\hbar}\right)^{4} \\
		&+ a_{6} \left(\frac{p}{\hbar}\right)^{6}
		+ a_{8} \left(\frac{p}{\hbar}\right)^{8}
		+ a_{10} \left(\frac{p}{\hbar}\right)^{10}.
    \end{aligned}
\end{equation}
Similarly, the first-order symmetry potential, defined as $U_{\mathrm{sym}}(\rho_0,p)$ $\equiv$ $\frac{\partial U_n(\rho_0,\delta,p)}{\partial\delta}\big|_{\delta=0}$ $=$ $-\frac{\partial U_p(\rho_0,\delta,p)}{\partial\delta}\big|_{\delta=0}$, can be also expressed as polynomial expansion on particle momentum~\cite{Wang:2024xzq},
\begin{equation}
	\begin{aligned}
		\label{eq:Usym_b}
		U_{\mathrm{sym}}(\rho_0,p) =~&b_{0}
		+ b_{2} \left(\frac{p}{\hbar}\right)^{2}
		+ b_{4} \left(\frac{p}{\hbar}\right)^{4} \\
		&+ b_{6} \left(\frac{p}{\hbar}\right)^{6}
		+ b_{8} \left(\frac{p}{\hbar}\right)^{8}
		+ b_{10} \left(\frac{p}{\hbar}\right)^{10}.
	\end{aligned}
\end{equation}
The above coefficients $a_{n}$ and $b_{n}$ ($n=0,2,4,6,8,10$) can be expressed in terms of the Skyrme parameters, and the details can be found in Ref.~\cite{Wang:2024xzq}.

An important quantity related to the single-nucleon potential is the nucleon effective mass.
The nucleon effective mass is used to characterize
the momentum dependence of the single-nucleon potential, and in nonrelativistic
models, it can be expressed as~\cite{Jaminon:1989wj,Li:2018lpy}
\begin{equation}
	\frac{m_{\tau}^{\ast}(\rho,\delta)}{m}=\left[1+\left.\frac{m}{p} \frac{\rm{d} U_\tau(\rho, \delta,p)}{\mathrm{d} {p}}\right|_{p=p_{F_{\tau}}}\right]^{-1},
\end{equation}
where $m$ is nucleon rest mass in vacuum and $p_{F_{\tau}}=\hbar(3 \pi^2 \rho_\tau)^{1/3}$ is the Fermi momentum of nucleons with isospin $\tau$.
The isoscalar nucleon effective mass $m_{s}^{\ast}$ is the nucleon effective mass in symmetric nuclear matter, and the isovector nucleon effective mass $m_{v}^{\ast}$ is the effective mass of proton (neutron) in pure neutron (proton) matter.
The nucleon effective mass splitting, denoted as $m_{n-p}^*(\rho, \delta) \equiv \left[m_{n}^{\ast}(\rho,\delta)-m_{p}^{\ast}(\rho,\delta) \right]/m$, is extensively used in nuclear physics.
$m_{n-p}^*(\rho, \delta)$ can be expanded as a power series in $\delta$, i.e.,
\begin{equation}
	\label{eq:spl_coes}
	m_{n-p}^*(\rho, \delta) =\sum_{n=1}^{\infty}
	\Delta m_{2 n-1}^{\ast}(\rho) \delta^{2 n-1},
\end{equation}
where $\Delta m_{2 n-1}^{\ast}(\rho)$ are the isospin splitting
coefficients (of the nucleon effective mass).
The first coefficient $\Delta m_{1}^{\ast}(\rho)$ is usually referred
to as the linear isospin splitting coefficient.

The nuclear matter EOS, conventionally defined as the binding energy per nucleon, can be derived from the Hamiltonian density as $E(\rho,\delta)$ = $\frac{{\cal H}(\mathbf{r})}{\rho}$.
For convenience, the density dependence of the symmetric part of the nuclear matter EOS and its isospin dependent part, i.e., the nuclear matter symmetry energy, are usually described by several characteristic quantities.
For the symmetric part of nuclear matter EOS in the N5LODD$4$ Skyrme pseudopotential used in this work, it is characterized by the following quantities:
the saturation density $\rho_{0}$, the binding energy per nucleon at saturation density $E_{0}(\rho_0)$, the incompressibility coefficient $K_{0}$, the skewness coefficient $J_{0}$, and the kurtosis coefficient $I_{0}$, all evaluated at $\rho_0$.
Specifically, these quantities are defined as:
\begin{equation}
K_{0}=\left. (3\rho_0)^2 \frac{\mathrm{d}^2 E_0(\rho)}{\mathrm{d} \rho^2}\right|_{\rho=\rho_0},
\end{equation}
\begin{equation}
J_{0}=\left.(3\rho_0)^3 \frac{\mathrm{d}^3 E_0(\rho)}{\mathrm{d} \rho^3}\right|_{\rho=\rho_0},
\end{equation}
\begin{equation}
I_{0}=\left.(3\rho_0)^4 \frac{\mathrm{d}^4 E_0(\rho)}{\mathrm{d} \rho^4}\right|_{\rho=\rho_0}.
\end{equation}
For the symmetry energy, we have similar characteristic quantities: $E_{\rm{sym}}(\rho_0)$
is the symmetry energy at $\rho_{0}$, $L$ is the slope parameter, $K_{\rm{sym}}$
is the curvature parameter, $J_{\rm{sym}}$ is the skewness parameters, and $I_{\rm{sym}}$ is the kurtosis parameters, of the symmetry energy, evaluated at $\rho_0$. These quantities are expressed as:
\begin{equation}
L=\left. 3\rho_0 \frac{\mathrm{d} E_{\rm{sym}}(\rho)}{\mathrm{d} \rho}\right|_{\rho=\rho_0},
\end{equation}
\begin{equation}
K_{\mathrm{sym}}  =\left.(3\rho_0)^2 \frac{\mathrm{d}^2 E_{\mathrm{sym}}(\rho)}{\mathrm{d} \rho^2}\right|_{\rho=\rho_0},
\end{equation}
\begin{equation}
J_{\mathrm{sym}}=\left.(3\rho_0)^3 \frac{\mathrm{d}^3 E_{\mathrm{sym}}(\rho)}{\mathrm{d} \rho^3}\right|_{\rho=\rho_0},
\end{equation}
\begin{equation}
I_{\mathrm{sym}}=\left.(3\rho_0)^4 \frac{\mathrm{d}^4 E_{\mathrm{sym}}(\rho)}{\mathrm{d} \rho^4}\right|_{\rho=\rho_0}.
\end{equation}

Based on the highly flexible N$5$LODD$4$ Skyrme pseudopotential~\cite{Wang:2024xzq} in terms of the density and momentum dependence, one can decouple the correlations between different characteristic quantities such as $m^*_{n-p}$, $K_0$, $J_0$ and $I_0$.
By implementing the N$5$LODD$4$ Skyrme pseudopotential into the BUU equation, we can study the individual influence of these quantities on specific observables like proton anisotropic flows in heavy-ion collisions.

\subsection{The lattice BUU transport model}
We employ the BUU equation~\cite{Bertsch:1988ik,BusPR512} to study proton anisotropic flows in the Au+Au collisions at $\sqrt{s_{\rm NN}}$ $=$ $2.4~\rm GeV$ as measured by the HADES
collaboration~\cite{HADES:2020lob,HADES:2022osk}.
The BUU equation deals with the time evolution of one-body phase-space distribution functions
(Wigner function) $f_\tau(\mathbf{r}, \mathbf{p}, t)$.
It reads
\begin{equation}
		(\partial_t + {\boldsymbol\nabla}_p \epsilon_\tau \cdot \boldsymbol\nabla_r - \boldsymbol\nabla_r \epsilon_\tau \cdot \boldsymbol\nabla_p) f_\tau = I^{\text{coll}}_\tau[f_n,f_p,\cdots],
\end{equation}
where $\tau$ represent different particle species.
The $\epsilon_\tau$ is the single-particle energy of the particle species $\tau$, and it contains the kinetic part and momentum-dependent mean-field potential $U_\tau(\mathbf{r}, \mathbf{p})$.
Note that $U_\tau(\mathbf{r}, \mathbf{p})$ should be treated as functionals of $f_\tau$.
In the present work, it is evaluated based on the Skyrme pseudopotential introduced in Sec.~\ref{Skyrme}.
The $I^{\text{coll}}_\tau$ is the collision integral, consisting of a gain term $(<)$ and a loss term $(>)$, i.e.,
\begin{equation}
	I^{\text{coll}}_\tau = \mathcal{K}^{<}_\tau[f_n,f_p,\cdots](1 \pm f_\tau) - \mathcal{K}^{>}_\tau[f_n, f_p,\cdots] f_\tau.
\end{equation}
The factor $1 \pm f_\tau$ in the gain term accounts for quantum statistics, with the plus sign for Bose enhancemen and the minus sign for Pauli blocking. The gain term and the loss term contain contributions from different scatterings.

In the present work, we include neutrons ($n$), protons ($p$), $\Delta$ and nucleon-resonances ($\Delta$, $N^*$, and $\Delta^*$), and $\pi$-mesons in the BUU equations.
The collision integral thus includes two-body scatterings, $NN \leftrightarrow N\Delta$, $\Delta(N^*,\Delta^*) \leftrightarrow N\pi$, and $\Delta^*(N^*) \leftrightarrow \Delta\pi$, whose scattering matrices can be deduced from their measured cross sections. We would like to note that in the present work, the light nuclei (e.g., deuteron, triton, $^{3}$He, $^{4}$He, etc.) are not included. The formation of light nuclei is an important aspect in heavy-ion collisions, and their treatment in the BUU transport model remains an intrinsically interesting and hot topic~\cite{Wang:2023gta,Wang:2025wim}. A detailed investigation of including explicitly light nuclei in the dynamical evolution and its influence on our present results will be pursued in future work.

The BUU equation is solved based on the test particle method~\cite{Wong:1982zzb}, where $f_\tau$ is mimicked by a large number of test particles, i.e.,
\begin{equation}
	f_\tau(\mathbf{r}, \mathbf{p}, t) = \frac{(2\pi\hbar)^3}{gN_E} \sum_{i\in\tau}^{A N_E} S\left[\mathbf{r}_i(t) - \mathbf{r}\right] \delta\left[\mathbf{p}_i(t) - \mathbf{p}\right].
\end{equation}
In the above equation, $A$ is the mass number of the system, $N_{\rm{E}}$ is the number of ensembles or number of test particles, and $S\left[\mathbf{r}_i(t) - \mathbf{r}\right]$ is a profile function in the coordinate space assigned to test particles to reduce numerical fluctuations.

In the LBUU transport model, we employ the lattice Hamiltonian method~\cite{Lenk:1989zz,Wang:2019ghr} to treat the mean-field evolution of the BUU equation, while the stochastic approach~\cite{Danielewicz:1991dh,Wang:2020xgk,Wang:2020ixf} to handle the collision integral $I_{\tau}^{\text{coll}}$.
The Thomas-Fermi initialization is applied to obtain the ground state of the nuclei (see Refs.~\cite{Wang:2019ghr,Wang:2020ixf} and the following section for details).
The Thomas-Fermi method actually provides a static solution of the BUU equation, and it therefore ensures the stability of the ground state evolution~\cite{Wang:2019ghr}.
The present framework for solving the BUU equation has been successfully applied to study nuclear collective motions in heavy nuclei~\cite{Wang:2019ghr,Wang:2020xgk,Wang:2020ixf,Song:2021hyw,Song:2023fnc,Song:2025bvt}, as well as light-nuclei yields~\cite{Wang:2023gta,Wang:2025wim}, and pion production~\cite{Li:2025uku} in heavy-ion collisions.

\begin{figure}[ht]
	\raggedright
	\includegraphics[width=\linewidth]{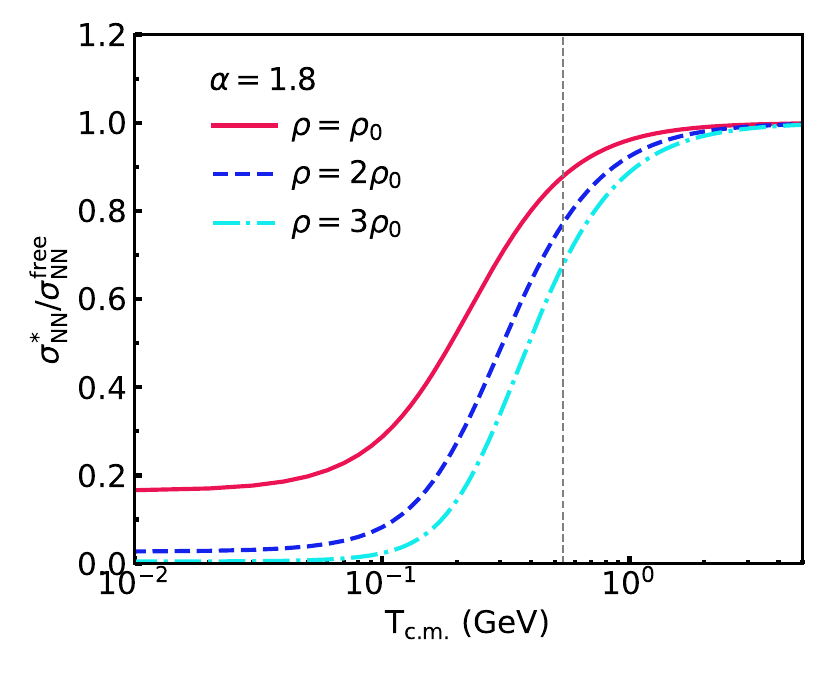}
	\caption{The total kinetic energy $T_{\rm{c.m.}}$ dependence of the medium correction for nucleon-nucleon elastic cross sections at different local density $\rho$ from Eq.~(\ref{CS-form}) with $\alpha = 1.8$. The vertical dashed line indicates the $T_{\rm{c.m.}}$ of two scattering nucleons corresponding to the HADES energy ($\sqrt{s_{\rm NN}}=2.4~\mathrm{GeV}$).}
	\label{fig:CS}
\end{figure}

In the present work, we perform the LBUU simulations of the Au+Au collisions at a beam energy of $1.23$~GeV/nucleon (corresponding
to $\sqrt{s_{\rm NN}} = 2.4 \, \text{GeV}$).
Some details of the simulations are as follows.
Protons are identified as free when their local densities are less than a critical density $\rho_{0}/8$~\cite{Li:2000bj,Chen:2002ym,Chen:2004si,Li:2005jy}, and we note that a $50\%$ variation of this critical density value has minor effects on the following results of proton anisotropic flows. The free nucleon-nucleon elastic cross sections $\sigma_{\rm NN}^{\text{free}}$ are adopted based on the parametrization of experimental nucleon-nucleon scattering data~\cite{Cugnon:1996kh}.
In order to investigate the in-medium effects, we also consider in-medium nucleon-nucleon elastic cross sections $\sigma_{\mathrm{NN}}^*$, which are parameterized based on thermodynamic $T$-matrix calculations~\cite{Alm:1994db,Alm:1995chb,Barker:2016hqv}, i.e.,
\begin{equation}
	\label{CS-form}
	\sigma_{\mathrm{NN}}^* = \sigma_{\mathrm{NN}}^{\mathrm{free}}
	\exp \bigg\{ -\alpha \frac{\rho / \rho_{0}}{1 + [T_{\mathrm{c.m.}}/(0.15~{\rm GeV})]^2}\bigg\} ,
\end{equation}
where the $\rho$ is local density of nucleons, the $T_{\rm{c.m.}}$ is total kinetic energy of two scattering nucleons at the rest frame of the local medium or cell.
The $\alpha$ are treated as a parameter that controls the strength of the in-medium effects, and it is chosen to be $1.8$, with which the width of isovector giant dipole resonance of \isotope[208]{Pb} can be reproduced~~\cite{Wang:2020xgk}.
The ratio $\sigma_{\mathrm{NN}}^*/\sigma_{\mathrm{NN}}^{\mathrm{free}}$ as a function of $T_{\mathrm{c.m.}}$ for various densities are shown in Fig.~\ref{fig:CS}.
The vertical black dashed line denotes the $T_{\rm{c.m.}}$ of two scattering nucleons corresponding to the HADES energy ($\sqrt{s_{\rm NN}}=2.4~\mathrm{GeV}$).
In Au+Au collisions at this energy, the $T_{\rm{c.m.}}$ of the majority of scattering nucleon pairs is expected to fall below this value.
Consequently, employing this in-medium cross section in the transport model may result in a significant in-medium suppression of the nucleon–nucleon cross sections during the collisions.

The nucleon resonances and $\Delta$ resonances are included up to $N(1720)$ and $\Delta(1950)$, respectively.
In the present work, a constant and isotropic elastic scattering cross section of $\sigma$ $=$ $40~\rm mb$ is adopted for two-body elastic scatterings between baryons other than nucleon-nucleon scatterings, following the standard practice in the transport model simulations~\cite{TMEP:2019yci,TMEP:2023ifw}. Nevertheless, we have verified that the flow observables discussed here are insensitive to this specific choice by varying this cross section by $\pm 50\%$.
For the $NN \rightarrow N\Delta$ scattering, a parameterized isotropic cross section as in Ref.~\cite{Li:2025uku} is employed.
For the $N\Delta \rightarrow NN$ scattering, its cross section is related to that of $NN \rightarrow N\Delta$ by the detailed balance condition (see~\cite{TMEP:2023ifw} for more details).
The resonance decay and productions, i.e., $\Delta(N^*,\Delta^*) \leftrightarrow N\pi$ and $\Delta^*(N^*) \leftrightarrow \Delta\pi$ are included with the cross sections and decay width taken from Refs.~\cite{TMEP:2023ifw,SMASH:2016zqf}.
When solving the BUU equation, we omit the single-particle potential of $\pi$ and all other resonances except $\Delta$. We note that neglecting the degrees of freedom of higher resonances ($N^*$ and $\Delta^*$) has a minor impact on flow observables in the present context.
It should be noted that the single-particle potentials of $\Delta$ are still very elusive~\cite{Drago:2014oja,Cozma:2014yna,Cai:2015hya,Li:2015hfa}.
In transport models of heavy-ion collisions, they are commonly given by liner combinations from those of $n$ and $p$.
In this work, we adopt the following $\Delta$ potentials:
\begin{eqnarray}
	 U_{\Delta^{++}} & = & -U_n + 2U_p,\nonumber\\
	U_{\Delta^{+}} & = & U_p,\nonumber\\
	U_{\Delta^{0}} & = & U_n,\nonumber\\
U_{\Delta^{-}} & = & 2U_n - U_p,
\end{eqnarray}
following Ref.~\cite{UmaMaheswari:1997mc}.
Note that the above $\Delta$ potentials do not cause extra differences in the potential energy between the initial and final states of the $NN$ $\leftrightarrow$ $N\Delta$ scatterings.

Furthermore, in the numerical aspect, the lattice spacing is set to be $1$ fm and the test particle number is taken to be $100,000$. The evolution of the reaction is terminated at $60~{\rm fm}/c$, with a time step of $0.2~{\rm fm}/c$.

\begin{figure*}[ht]
	\centering
	\includegraphics[width=\linewidth]{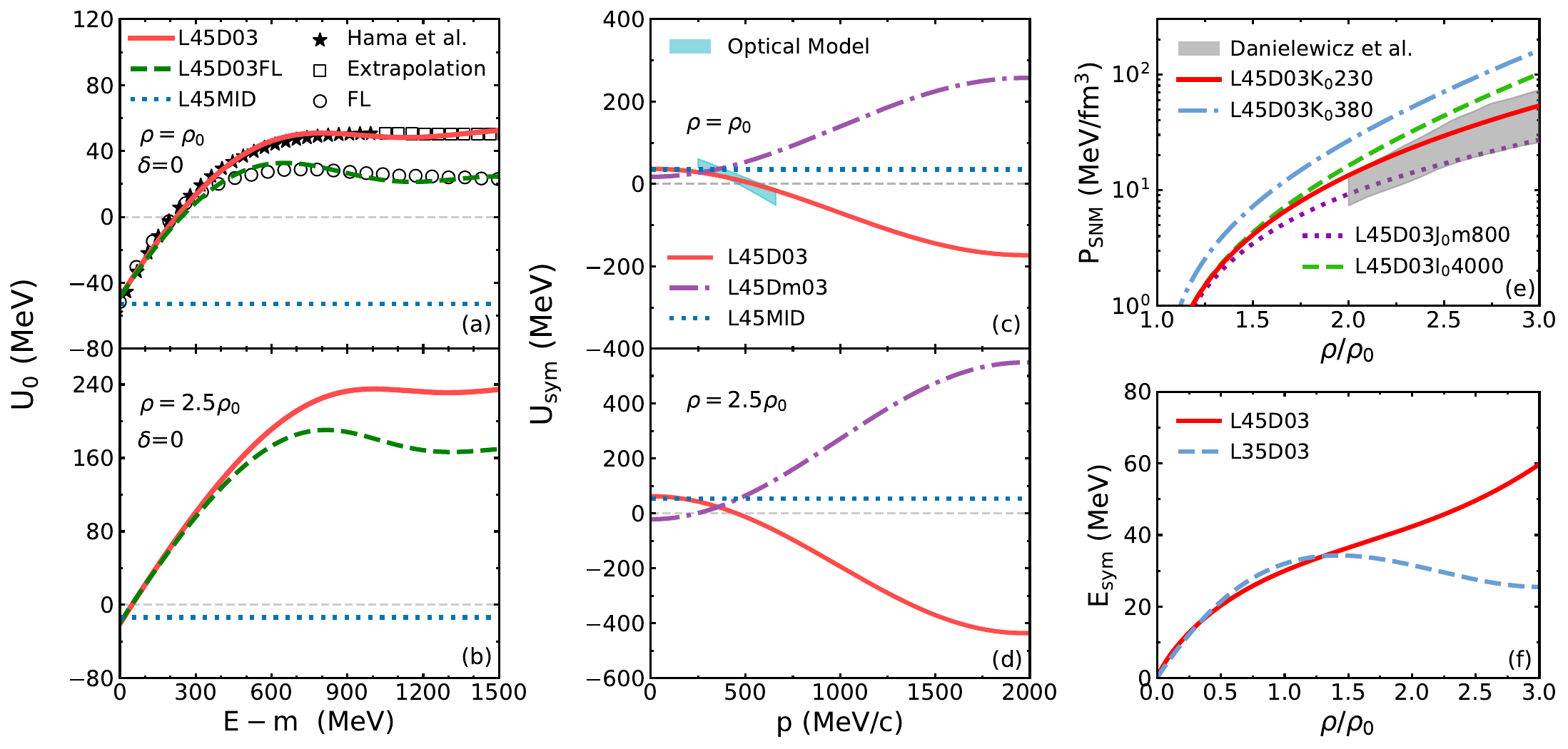}
	\caption{Panel~(a): The kinetic energy dependence of the single-nucleon potential in cold symmetric nuclear matter predicted by the interactions L$45$D$03$, L$45$D$03$FL and L$45$MID. The nucleon optical potential (Schrödinger equivalent potential) in symmetric nuclear matter at saturation density $\rho_{0}$ obtained by Hama $et$ $al.$~\cite{Hama:1990vr} is shown as black stars. The black squares show the extrapolation of Hama’s data. The black circles named FL are the optical potential obtained by Feldmeier and Lindner in Ref.~\cite{Feldmeier:1991ey}. Panel~(b): Same as Panel (a) but for the density $\rho=2.5\rho_{0}$.
Panel~(c): The momentum dependence of the symmetry potential $U_{\rm sym}(\rho_{0},p)$ given by the interactions with distinct $U_{\rm sym}(\rho_{0},p)$, namely, L$45$D$03$, L$45$Dm$03$ and L$45$MID. The results from the global optical model analyses~\cite{Li:2014qta,Xu:2010fh} are shown in light blue band.
Panel~(d): Same as Panel (c) but for the density $\rho=2.5\rho_{0}$.
Panel~(e): The pressure of symmetric nuclear matter ($P_{\text{SNM}}$) as a function of nucleon density given by the interactions: L$45$D$03$, L$45$D$03$K$_{0}380$, L45D$03$J$_{0}$m$800$ and L45D$03$I$_{0}$$4000$. The gray band shows the constrains obtained by Danielewize et al.~\cite{Danielewicz:2002pu}.
Panel~(f): The density dependence of the symmetry energy ($E_{\rm sym}$) for interactions L45D03 and L35D03.}
	\label{fig:U0Usy}
\end{figure*}

\subsection{Description of the interactions employed}
In the present study, we employ several Skyrme pseudopotentials derived from the N$5$LODD$4$ formulation.
In this formulation, there are $30$ Skyrme parameters, which can be mapped to $30$ macroscopic quantities, namely $10$ characteristic quantities ($\rho_0$, $E_0(\rho_0)$, $K_0$, $J_0$, $I_0$, $E_{{\mathrm{sym}}}(\rho_0)$, $L$, $K_{{\mathrm{sym}}}$, $J_{{\mathrm{sym}}}$ and $I_{{\mathrm{sym}}}$) for the EOS of symmetric nuclear matter and the symmetry energy, $10$ coefficients $a_{n}$ and $b_{n}$ ($n = 2,4,6,8,10$) for single particle potential in Eqs.~(\ref{eq:U0_N5LO}) and~(\ref{eq:Usym_b}), and $10$ gradient term parameters $E^{[n]}$ and $F^{[n]}$ ($n = 2,4,6,8,10$)~\cite{Wang:2024xzq}.
We first choose the desired values of these macroscopic quantities, and the Skyrme parameters can then be determined accordingly.
These parameters are summarized in Table~\ref{tab:Paras_Quants}.
The baseline interaction SP10DD4L45D03 is constructed following the N$5$LO Skyrme pseudopotential SP$10$L$45$D$03$ in Ref.~\cite{Wang:2024xzq}.
For the baseline interaction SP$10$DD$4$L$45$D$03$, the prefix “SP10” indicates the momentum-derivative terms in the Skyrme pseudopotential are taken up to the tenth order, and “DD4” denotes the inclusion of density-dependent terms up to the fourth order (see Sec.~\ref{Skyrme} for more details ). The label "L$45$" refers to the slope parameter of the symmetry energy (L=$45$ MeV). "D$03$" indicates the linear isospin splitting coefficient of the nucleon effective mass at saturation density, $\Delta m_{1}^{\ast}(\rho_0)$ $=$ $0.3$, which is supported by global optical-model analyses~\cite{Li:2014qta,Xu:2010fh}. The baseline interaction SP$10$DD$4$L$45$D$03$ serves as a reasonable physics-wise choice as it is capable of reproducing a number of experimental and theoretical constraints~\cite{Wang:2024xzq}: (i) the energy dependence of the Hama potential and its extrapolation up to $1.5$ GeV~\cite{Hama:1990vr,Cooper:1993nx} (see also the results shown later); (ii) key astrophysical observables such as the mass-radius relation and the tidal deformability of neutron stars~\cite{Wang:2024xzq}; (iii) the empirical nuclear saturation properties ($\rho_{0}=0.16$ $\rm{fm}^{-3}$,$E_{0}=-16$ MeV and $K_{0}=230$ MeV) from nuclear structure data; (iv) the constraints from flow data in HICs~\cite{Danielewicz:2002pu} (see also the results shown later). The labels in other interactions indicate specific variations relative to this baseline, which allows us to clearly isolate the influence of individual physical ingredients on the flow observables.
We then employ different interactions within the lattice BUU transport model to investigate the sensitivity of the proton anisotropic flows to various quantities.
We first illustrate different behaviors of these interactions in terms of the single-particle potential and the nuclear matter EOS. In the following, we omit the prefix SP$10$DD$4$ of these interactions for clarity.
\begin{table*}
	\caption{
		The $30$ independent model parameters and the $30$ adjustable macroscopic quantities for the $\mathrm{N5LODD4}$ Skyrme pseudopotential.
	}
	\label{tab:Paras_Quants}
	\begin{tabular}{|c|c|}
		\hline
		Parameters & Quantities \\ \hline
		$t_0$, $x_0$, $t_{3}^{[1]}$, $x_{3}^{[1]}$,$t_3^{[3]}$, $x_{3}^{[3]}$,
		$t_3^{[5]}$, $x_{3}^{[5]}$, $t_3^{[7]}$, $x_{3}^{[7]}$, &
		$\rho_0$, $E_0(\rho_0)$, $K_0$, $J_0$, $I_0$, $E_{{\mathrm{sym}}}(\rho_0)$, $L$, $K_{{\mathrm{sym}}}$, $J_{{\mathrm{sym}}}$, $I_{{\mathrm{sym}}}$, \\ \hline
		$t_{1}^{[2]}$, $t_{2}^{[2]}$, $t_{1}^{[4]}$, $t_{2}^{[4]}$,
		$t_{1}^{[6]}$, $t_{2}^{[6]}$, $t_{1}^{[8]}$, $t_{2}^{[6]}$,
		$t_{1}^{[10]}$, $t_{2}^{[10]}$, &
		$a_2$, $a_4$, $a_6$, $a_8$, $a_{10}$, $b_2$, $b_4$, $b_6$, $b_8$, $b_{10}$,  \\ \hline
		$x_{1}^{[2]}$, $x_{2}^{[2]}$, $x_{1}^{[4]}$, $x_{2}^{[4]}$,
		$x_{1}^{[6]}$, $x_{2}^{[6]}$, $x_{1}^{[8]}$, $x_{2}^{[8]}$,
		$x_{1}^{[10]}$, $x_{2}^{[10]}$ &
		$E^{[2]}$, $F^{[2]}$, $E^{[4]}$, $F^{[4]}$, $E^{[6]}$, $F^{[6]}$, $E^{[8]}$, $F^{[8]}$, $E^{[10]}$, $F^{[10]}$ \\ \hline
	\end{tabular}
\end{table*}

Based on the baseline interaction L45D03, the interaction L$45$D$03$FL is constructed to reproduce the optical potential obtained by Feldmeier and Lindner~\cite{Feldmeier:1991ey}, rather than the Schr\"{o}dinger equivalent potential obtained by Hama potential~\cite{Hama:1990vr,Cooper:1993nx} (Note that both of them are based on the same proton scattering data in Ref.~\cite{Hama:1990vr} but with different non-relativistic reductions) used for constructing L$45$D$03$.
This is done by readjusting the parameter $a_{n}$ in Eq.~(\ref{eq:U0_N5LO}).
The single-particle potential $U_{0}$ as a function of the nucleon kinetic energy $E-m$ $=$ $\sqrt{p^2+m^2}+U_{0}(\rho_{0},p)-m$ in cold symmetric nuclear matter at $\rho_0$, for the L45D03 and L45D03FL, are presented in panel (a) of Fig.~\ref{fig:U0Usy}, and compared with Hama potential~\cite{Hama:1990vr} or the one obtained by Feldmeier and Lindner~\cite{Feldmeier:1991ey}.

Considering the significant uncertainty in the symmetry potential, an interaction L$45$Dm$03$ with distinct isospin splitting coefficient at saturation density $\Delta m_{1}^{\ast}(\rho_0)$ $=$ $-0.3$~\cite{Wang:2024xzq} contrast to that of L$45$D$03$ ($\Delta m_{1}^{\ast}(\rho_0)$ $=$ $0.3$) is also adopted.
It is obtained by adjusting the $b_{n}$ parameters in Eq.~(\ref{eq:Usym_b})~\cite{Wang:2024xzq}.
The symmetry potential as a function of momentum $U_{\mathrm{sym}}(\rho_{0},p)$ in cold symmetric nuclear matter at $\rho_0$ for L$45$D$03$ and L$45$Dm$03$ are shown in panel (c) of Fig.~\ref{fig:U0Usy}.
Also shown is the momentum dependence of $U_{\mathrm{sym}}(\rho_0,p)$ obtained from a global optical model analyses~\cite{Li:2014qta,Xu:2010fh}, and the value of $\Delta m_{1}^{\ast}(\rho_0)$ is predicted to be $(0.41\pm0.15)$.

Additionally, a momentum-independent interaction, L$45$MID is adopted to further explore the role of the momentum dependence in the nucleon single-particle potential.
The $U_{0}$ and $U_{\mathrm{sym}}$ of L$45$MID for cold symmetric nuclear matter at $\rho_0$ are constant values of $-52.84~\rm MeV$ and $35.43~\rm MeV$, as shown in panels (a) and (c) in Fig.~\ref{fig:U0Usy}, respectively.
To further illustrate the density dependence
of the single-nucleon potential, panels (b) and (d) of Fig.~\ref{fig:U0Usy} present the corresponding optical potentials and symmetry potentials at $2.5\rho_0$.

Furthermore, the interactions with different values of $K_0=380$ MeV, $J_0=-800$ MeV and $I_0=4000$ MeV relative to those of L$45$D$03$, are also constructed and labeled as K$_0380$, J$_0$m$800$, and I$_04000$, respectively. The corresponding pressure $P_{\rm SNM}$ as a function of density for symmetric nuclear matter of the above interactions are depicted in panel (e) of Fig.~\ref{fig:U0Usy}. The EOS of the default interaction L$45$D$03$ is compared with those of interactions featuring a stiffer incompressibility coefficient (L$45$D$03$K$_{0}$$380$), a modified skewness coefficient $J_0$ (L$45$D$03$J$_{0}$m$800$), and a modified kurtosis coefficient $I_0$ (L$45$D$03$I$_{0}$$4000$). The gray band represents the constraints from flow data in heavy-ion collisions~\cite{Danielewicz:2002pu}.

To explore the influence of the symmetry energy, particularly its behavior at high densities, we constructed the interaction labeled L35D03 by modifying the L45D03 parameter set with the following values: $E_{\text{sym}}(\rho_0)=32$ MeV, $L=35$ MeV, $K_{{\mathrm{sym}}}=-300$ MeV, $J_{{\mathrm{sym}}}=720$ MeV, and $I_{{\mathrm{sym}}}=622.1$ MeV, as provided in Ref.~\cite{Ye:2024meg}. This specific parameter choice results in a softer symmetry energy in the density range of around $1.5\rho_{0}$-$3\rho_{0}$, which is relevant for addressing issues such as the hyperon puzzle, as detailed therein~\cite{Ye:2024meg}. The corresponding density dependence of the symmetry energy ($E_{\rm sym}$) for interactions L$45$D$03$ and L$35$D$03$ are shown in panel (f) of Fig.~\ref{fig:U0Usy}.

Meanwhile, to explore the effect of the in-medium correction of the nucleon-nucleon elastic cross sections, we introduce the interaction L$45$D$03$ with CS, which shares the parameters of L$45$D$03$ but includes an in-medium correction using the parametrization form from Eq.~(\ref{CS-form})~\cite{Wang:2020xgk}.

In Table~\ref{tab:SPpara}, we present the macroscopic characteristic quantities (i.e., $\rho_0$, $E_0(\rho_0)$, $K_0$, $J_0$, $I_0$, $E_{{\mathrm{sym}}}(\rho_0)$, $L$, $K_{{\mathrm{sym}}}$, $J_{{\mathrm{sym}}}$, $I_{{\mathrm{sym}}}$, $\Delta m_{1}^*(\rho_0)$, $a_{0}$, $a_{2}$, $a_{4}$, $a_{6}$, $a_{8}$, $a_{10}$, $b_{0}$, $b_{2}$, $b_{4}$, $b_{6}$, $b_{8}$, $b_{10}$ and $E^{[2]}$) with the above interactions.
Here, for simplicity, all gradient parameters except $E^{[2]}$ are set to be $0$, and thus we have totally $21$ adjustable quantities as shown in the right column of Table~\ref{tab:Paras_Quants}.
The $E^{[2]}$ parameter, as described in Ref.~\cite{Wang:2019ghr,Wang:2024xzq}, is modified for each interaction to reproduce the experimental ground state binding energy of $^{197}$Au within the Thomas-Fermi method.
Also listed in Table~\ref{tab:SPpara} are the binding energy ($E_{\rm B}$), proton root-mean-square radius ($\sqrt{\langle r^2 \rangle_p}$), neutron root-mean-square radius ($\sqrt{\langle r^2 \rangle_n}$), nucleon root-mean-square radius $\sqrt{\langle r^2 \rangle_N}$ and neutron skin thickness $\Delta r_{\rm{np}}$ of $^{197}$Au for all the interactions in Thomas-Fermi initialization.
For comparison, we also list the experimental binding energy~\cite{Wang:2021xhn} and the empirical point-proton root-mean-square radius of $^{197}$Au. It should be noted that the empirical value of the point-proton root-mean-square radius is obtained from the experimental charge radius of Ref.~\cite{Angeli:2013epw} using the corrections from Ref.~\cite{Ong:2010gf} by neglecting the spin-orbit contribution.
It is seen from Table~\ref{tab:SPpara} that, for each interaction, the Thomas-Fermi method can reasonably reproduce the empirical value of the point-proton root-mean-square radius.

\begin{table*}[ht]
	\label{tab:SP}
	\caption{Macroscopic characteristic quantities of eight interactions: L$45$D$03$, L$45$D$03$FL, L$45$Dm$03$, L$45$MID, L$45$D$03$K$_{0}380$, L$45$D$03$J$_{0}$m$800$, L$45$D$03$I$_{0}4000$, and L$35$D$03$. The obtained binding energy $E_{\rm B}$, proton root-mean-square radius $\sqrt{\langle r^2 \rangle_p}$, neutron root-mean-square radius $\sqrt{\langle r^2 \rangle_n}$, nucleon root-mean-square radius $\sqrt{\langle r^2 \rangle_N}$ and neutron skin thickness $\Delta r_{\rm{np}}$ of $^{197}$Au in Thomas-Fermi initialization are also shown. Also listed are the experimental binding energy of $^{197}$Au~\cite{Wang:2021xhn}, along with its point-proton root-mean-square radius, which is obtained from the experimental charge radius of Ref.~\cite{Angeli:2013epw} using the corrections without including spin-orbit contribution from Ref.~\cite{Ong:2010gf}. }
	\label{tab:SPpara}
	\centering
	\fontsize{6.0}{10}\selectfont
	\setlength{\tabcolsep}{1pt} 
	\begin{tabular}{ccccccccccccc}
		\toprule
		\hline
		& L45D03 & L45D03FL & L45Dm03 & L45MID & L45D03K$_{0}$380 & L45D03J$_{0}$m800 & L45D03I$_{0}$4000 & L35D03 & Expt. \\
		\midrule
		$\rho_{0}$ ($\rm{fm}^{-3}$) & 0.160 & 0.160 & 0.160 & 0.160 & 0.160 & 0.160 & 0.160 & 0.160 & -\\
		$E_0(\rho_0)$ (MeV) & -16.0 & -16.0 & -16.0 & -16.0 & -16.0 & -16.0 & -16.0 & -16.0 & -\\
		$K_0$ (MeV) & 230.0 & 230.0 & 230.0 & 230.0 & 380.0 & 230.0 & 230.0 & 230.0 & -\\
		$J_0$ (MeV) & -383.0 & -383.0 & -383.0 & -383.0 & -383.0 & -800.0 & -383.0 & -383.0 & -\\
		$I_0$ (MeV) & 1818.9 & 1818.9 & 1818.9 & 1818.9 & 1818.9 & 1818.9 & 4000 & 1818.9 & -\\
		$E_{\text{sym}}(\rho_0)$ (MeV) & 30 & 30 & 30 & 30 & 30 & 30 & 30 & 32 & -\\
		$L$ (MeV) & 45 & 45 & 45 & 45 & 45 & 45 & 45 & 35 & -\\
		$K_{\rm sym}$ (MeV) & -110 & -110 & -110 & -110 & -110 & -110 & -110 & -300 & -\\
		$J_{\rm sym}$ (MeV) & 700 & 700 & 700 & 700 & 700 & 700 & 700 & 720 & -\\
		$I_{\rm sym}$ (MeV) & -2458.5 & -2458.5 & -2458.5 & -2458.5 & -2458.5 & -2458.5 & -2458.5 & 622.1 & -\\
		$\Delta m_{1}^*(\rho_0)$ & 0.3 & 0.3 & -0.3 & 0 & 0.3 & 0.3 & 0.3 & 0.3 & -\\
		$a_0$ (MeV) & -64.97 & -64.44 & -64.97 & -52.84 & -64.97 & -64.97 & -64.97 & -64.97 & -\\
		$a_2$ (MeV fm$^2$) & 7.104 & 6.829 & 7.104 & 0 & 7.104 & 7.104 & 7.104 & 7.104 & -\\
		$a_4$ (MeV fm$^4$) & -0.1628 & -0.1719 & -0.1628 & 0 & -0.1628 & -0.1628 & -0.1628 & -0.1628 & -\\
		$a_6$ (MeV fm$^6$) & $1.731{\times}10^{-3}$ & $1.908{\times}10^{-3}$ & $1.731{\times}10^{-3}$ & 0 & $1.731{\times}10^{-3}$ & $1.731{\times}10^{-3}$ & $1.731{\times}10^{-3}$ & $1.731{\times}10^{-3}$ & -\\
		$a_8$ (MeV fm$^8$) & $-8.614{\times}10^{-6}$ & $-9.672{\times}10^{-6}$ & $-8.614{\times}10^{-6}$ & 0 & $-8.614{\times}10^{-6}$ & $-8.614{\times}10^{-6}$ & $-8.614{\times}10^{-6}$ & $-8.614{\times}10^{-6}$ & -\\
		$a_{10}$ (MeV fm$^{10}$) & $1.621{\times}10^{-8}$ & $1.827{\times}10^{-8}$ & $1.621{\times}10^{-8}$ & 0 & $1.621{\times}10^{-8}$ & $1.621{\times}10^{-8}$ & $1.621{\times}10^{-8}$  & $1.621{\times}10^{-8}$& -\\
		$b_0$ (MeV) & 42.73 & 42.73 & 23.32 & 35.43 & 42.73 & 42.73 & 42.73 & 42.73 & -\\
		$b_2$ (MeV fm$^2$) & -5.168 & -5.168 & 5.918 & 0 & -5.168 & -5.168 & -5.168 & -5.168 \\
		$b_4$ (MeV fm$^4$) & $4.250{\times}10^{-2}$ & $4.2503{\times}10^{-2}$ & $-4.867{\times}10^{-2}$ & 0 & $4.250{\times}10^{-2}$ & $4.250{\times}10^{-2}$ & $4.250{\times}10^{-2}$ & $4.250{\times}10^{-2}$ & -\\
		$b_6$ (MeV fm$^6$) & $-1.398{\times}10^{-4}$ & $-1.398{\times}10^{-4}$ & $1.601{\times}10^{-4}$ & 0 & $-1.398{\times}10^{-4}$ &  $-1.398{\times}10^{-4}$ & $-1.398{\times}10^{-4}$ & $-1.398{\times}10^{-4}$ & -\\
		$b_8$ (MeV fm$^8$) & $2.4645{\times}10^{-7}$ & $2.4645{\times}10^{-7}$ & $-2.822{\times}10^{-7}$ & 0 & $2.4645{\times}10^{-7}$ & $2.4645{\times}10^{-7}$ & $2.4645{\times}10^{-7}$ & $2.4645{\times}10^{-7}$ & -\\
		$b_{10}$ (MeV fm$^{10}$) & $-2.7026{\times}10^{-10}$ & $-2.7026{\times}10^{-10}$ & $3.095{\times}10^{-10}$ & 0 & $-2.7026{\times}10^{-10}$ & $-2.7026{\times}10^{-10}$ & $-2.7026{\times}10^{-10}$ & $-2.7026{\times}10^{-10}$ & -\\
		$E^{[2]}$ (${\rm MeV\cdot fm}^5$) & $-310$ & $-310$ & $-310$ & $-310$ & $-200$ & $-285$ & $-310$ & $-305$ & -\\
		$E_{\rm B}$ (${\rm MeV}$) & $-1559.95$ & $-1559.95$ & $-1559.94$ & $-1560.10$ & $-1560.01$ & $-1558.54$ & $-1556.90$ & $-1557.35$ & $-1559.37$~\cite{Wang:2021xhn}\\
		$\sqrt{\langle r^2 \rangle_p} \, (\rm{fm})$  & $5.443$ & $5.443$ & $5.443$ & $5.443$ & $5.274$ & $5.406$ & $5.436$ & $5.418$ & $5.371$~\cite{Ong:2010gf,Angeli:2013epw}\\
		$\sqrt{\langle r^2 \rangle_n} \, (\rm{fm})$  & $5.556$ & $5.556$ & $5.556$ & $5.556$ & $5.345$ & $5.509$ & $5.548$ & $5.569$ & -\\
		$\sqrt{\langle r^2 \rangle_N} \, (\rm{fm})$  & $5.511$ & $5.511$ & $5.511$ & $5.511$ & $5.318$ & $5.468$ & $5.503$ & $5.501$ & -\\
		$\Delta r_{\rm{np}} \, (\rm{fm})$  & $0.113$ & $0.113$ & $0.113$ & $0.113$ & $0.070$ & $0.102$ & $0.111$ & $0.151$ & -\\
		\hline
		\bottomrule
	\end{tabular}
\end{table*}

We would like to point out that the variations of the macroscopic quantities in the interactions constructed above roughly reflect the current uncertainties of these macroscopic quantities.
Given this context, we present the LBUU transport model predictions for the directed ($v_1$), elliptic ($v_2$), triangular ($v_3$), and quadrangular ($v_4$) flows of free protons in mid-peripheral ($20$–$30\%$ centrality) Au+Au collisions at $\sqrt{s_{\rm NN}}=2.4~\mathrm{GeV}$.
These transport model predictions, which employ the various interactions described above, are used to investigate the effects of several key physical inputs, including the momentum dependence of the nucleon single-particle potential, the stiffness of symmetric nuclear matter EOS, the symmetry energy, and the in-medium correction to nucleon-nucleon elastic cross sections. Finally, we systematically benchmark our theoretical results against the collective flow data measured by the HADES collaboration~\cite{HADES:2020lob,HADES:2022osk}.

\subsection{Anisotropic flows in heavy-ion collisions and centrality selection}

The anisotropic flow coefficients $v_n$ quantify the anisotropy and are defined as the Fourier components in the expansion of the particle transverse momentum $p_t$ spectra with respect to the azimuthal angle $\phi$ relative to the reaction plane~\cite{Poskanzer:1998yz}:
\begin{equation}
\label{eq:flow_n}
\small
E \frac{\mathrm{d}^3 N}{\mathrm{d} p^3}= \frac{1}{2\pi}\frac{\mathrm{d}^{2} N}{p_{t} \mathrm{d} p_{t} \mathrm{d} y } \left[ 1 + \sum_{n=1}^{\infty} 2 v_{n} (p_{t},y) \cos{(n \phi)} \right].
\end{equation}
The flow coefficients $v_n$ generally depend on both the transverse momentum $p_t$ and rapidity $y$ of the particles. For a given $y$, the coefficients at a specific $p_t$ can be evaluated as
\begin{equation}
\label{eq:flow_average}
v_{n} (p_{t}) = \left \langle \cos{(n \phi)} \right \rangle,
\end{equation}
where $\langle \cdots \rangle$ denotes an average over the azimuthal distribution of particles with transverse momentum $p_t$.
Furthermore, the anisotropic flow coefficients $v_n$ can be expressed as particle number averages in terms of single-particle momentum components~\cite{Chen:2004dv,Chen:2004vha}:
\begin{align}
\label{eq:flow_expressions}
v_{1}(p_{t}) &= \left \langle \frac{p_{x}}{p_{t}} \right \rangle, \\
v_{2}(p_{t}) &= \left \langle \frac{p_{x}^{2}-p_{y}^{2}}{p_{t}^{2}} \right \rangle, \\
v_{3}(p_{t}) &= \left \langle \frac{p_{x}^{3}- 3 p_{x} p_{y}^{2}}{p_{t}^{3}} \right \rangle, \\
v_{4}(p_{t}) &= \left \langle \frac{p_{x}^{4}- 6 p_{x}^{2} p_{y}^{2} + p_{y}^{4} }{p_{t}^{4}} \right \rangle,
\end{align}
where $p_x$ and $p_y$ are the components of the particle momentum parallel and perpendicular to the reaction plane, respectively.

The anisotropic flows originate from the spatial anisotropy of the participant matter of the collision and are therefore highly sensitive to the geometry of the overlap region of the colliding nuclei or, equivalently, to the collision centrality.
Experimentally, centrality classes are conventionally inferred via Glauber model Monte Carlo simulations, which map detector-measured observables—such as charged particle multiplicity to estimates of impact parameter and the number of participant nucleons $N_{\rm{part}}$ through statistical averaging over collision ensembles.  Theoretically, centrality is typically quantified by the impact parameter.
This geometric parameter is mapped to experimentally accessible observables via Glauber model Monte Carlo calculations.
However, the mapping from observables to centrality classes involves significant model dependence and remains a significant challenge in heavy-ion collision studies. To facilitate direct comparisons between theoretical predictions and experimental data, a practical approach is to employ a fixed impact parameter or predefined ranges derived from the Glauber model~\cite{HADES:2017def}. In this work, we implement a simplified strategy wherein each centrality class is mapped to a representative range of impact parameter, as quantified by Glauber model calculations for the HADES experiment~\cite{HADES:2017def}.
In the centrality range of 20–30\%, which is the main focus of the present work, the corresponding impact parameter ranges from $6.6$ to $8.1~\rm fm$~\cite{HADES:2017def}.
However, realistic impact parameter distributions for a given centrality class are characterized by quasi-Gaussian profiles rather than discrete values. To account for this, we perform simulations at multiple discrete impact parameters that collectively sample the full distribution predicted by the Glauber model. The final observables are then calculated through a weighted superposition of these individual simulations, with weighting factors determined by the probability distribution of impact parameters within each centrality class. This approach of determining the centrality is not exactly the same as the experimental method but it ensures a more systematic and statistically meaningful comparison between theoretical predictions and experimental observations.

\section{Results and discussion}
\label{discussion}
In this section, we first examine the evolution of the $^{197}\rm{Au}$ ground state and the centre density of the collision system for Au+Au collisions at HADES energy.
Then, we present the results of $v_1$, $v_2$, $v_3$, and $v_4$ for free protons in mid-peripheral (20-30$\%$ centrality class) Au+Au collisions at $\sqrt{s_{\rm NN}}=2.4\,\mathrm{GeV}$, as predicted by the LBUU transport model using the various interactions described above, to demonstrate the effects of the momentum dependence of nucleon mean field potential including the symmetry potential, the stiffness of symmetric nuclear matter EOS, the symmetry energy and the in-medium correction of nucleon-nucleon elastic cross sections. These results are then compared with the corresponding proton collective flow data from the HADES collaboration~\cite{HADES:2020lob,HADES:2022osk}.

\subsection{Evolutions of ground state properties and the centre density of collision system}
\begin{figure}[ht]
	\raggedright
	\includegraphics[width=\linewidth]{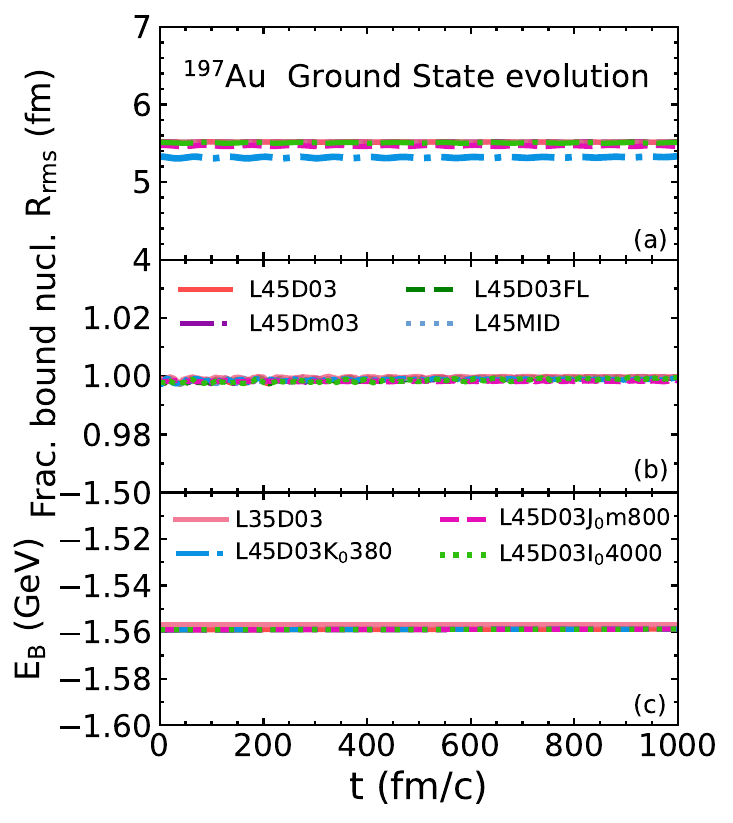}
	\caption{Time evolution of (a) nucleon root-mean-square radius, (b) the fraction of bound nucleons, and (c) binding energy $E_{\rm B}$ of ground state $^{197}$Au with interactions: L$45$D$03$, L$45$D$03$FL, L$45$Dm$03$, L$45$MID, L$35$D$03$,  L$45$D$03$K$_{0}380$, L$45$D$03$J$_{0}$m$800$ and L$45$D$03$I$_{0}4000$ up to $1000~{\rm fm}/c$. Calculations are performed with time step $\Delta t = 0.2~{\rm fm}/c$ and $100,000$ test particles. }
	\label{fig:GSeve}
\end{figure}

In this subsection, we first examine the evolution of the $^{197}\rm{Au}$ ground state. Following the Thomas-Fermi initialization, we calculate the mean-field evolution (note that the collision term should vanish due to Pauli blocking) of a ground-state \isotope[197]{Au} within the LBUU transport model for the interactions: L$45$D$03$, L$45$D$03$FL, L$45$Dm$03$, L$45$MID, L$35$D$03$, L$45$D$03$K$_{0}380$, L$45$D$03$J$_{0}$m$800$ and L$45$D$03$I$_{0}4000$.
Fig.~\ref{fig:GSeve} show the time evolution of (a) the nucleon root-mean-square radius, (b) the fraction of bound nucleons, and (c) the binding energy $E_{\rm B}$.
The details of calculating the fraction of bound nucleons can be found in Ref.~\cite{Wang:2019ghr}.
It is clearly seen from Fig.~\ref{fig:GSeve} that the ground-state nucleus maintains very good stability for all interactions investigated, with the corresponding values very accurately preserved throughout the time evolution.

\begin{figure}[ht]
	\raggedright
	\includegraphics[width=\linewidth]{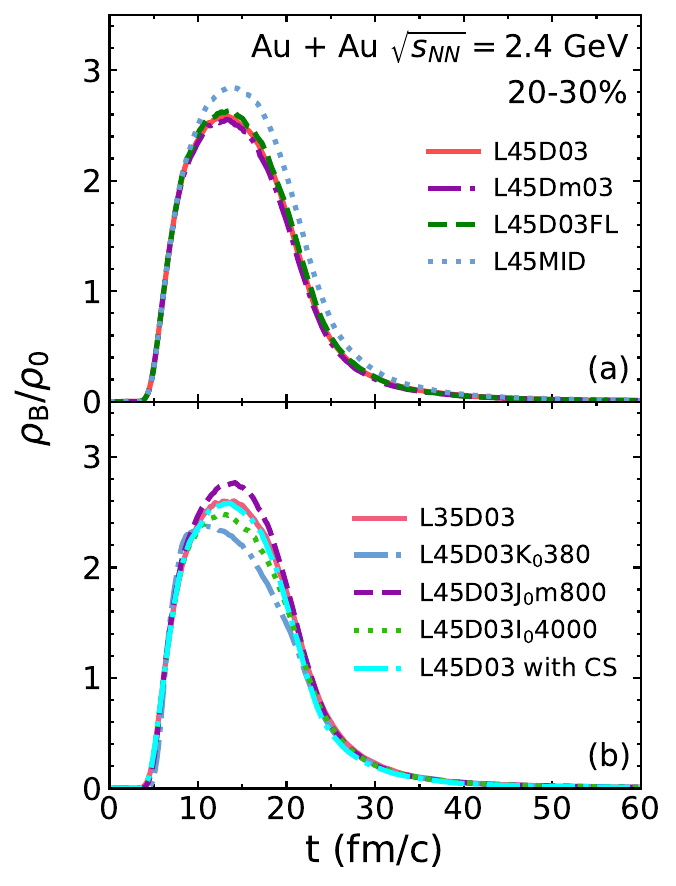}
	\caption{Time evolution of the central baryon density $\rho_{\rm{B}}$ in Au+Au collisions at $\sqrt{s_{\rm NN}}=2.4~\mathrm{GeV}$ and $20$--$30\%$ centrality predicted by the LBUU transport model with interactions (a) L$45$D$03$, L$45$D$03$FL, L$45$Dm$03$ and L$45$MID as well as (b) L$35$D$03$,  L$45$D$03$K$_{0}380$, L$45$D$03$J$_{0}$m$800$, L$45$D$03$I$_{0}4000$ and L$45$D$03$ with CS.}
	\label{fig:cenrho}
\end{figure}

In addition, the time evolution of the density in a collision system represents a crucial aspect for probing the collision dynamics.
We analyze here the time evolution of the baryon density $\rho_{\rm{B}}$ in the center of the reaction zone in mid-central ($20$--$30\%$ centrality) Au+Au collisions at $\sqrt{s_{\rm NN}}=2.4~\mathrm{GeV}$.
In particualr, Fig.~\ref{fig:cenrho} displays the time evolution of the baryon density $\rho_{\rm{B}}$ in the central region (a $1 \times 1 \times 1~\mathrm{fm}^{3}$ cell) obtained from the LBUU simulations using interactions: L$45$D$03$, L$45$D$03$FL, L$45$Dm$03$, L$45$MID, L$35$D$03$,  L$45$D$03$K$_{0}380$, L$45$D$03$J$_{0}$m$800$, L$45$D$03$I$_{0}4000$ and L$45$D$03$ with CS. These results provide insight into how the choice of interaction influences the dynamical behaviors of high-density nuclear matter created during the collision. One can see that the maximum density reaches about $2.8\rho_0$ for the momentum-independent interaction L$45$MID and the interaction L$45$D$03$J$_{0}$m$800$ with quite soft SNM EOS from a small $J_0$ value of $-800$~MeV (see Fig.~\ref{fig:U0Usy}~(e)), while it becomes to about $2.4\rho_0$ for the interaction L$45$D$03$K$_{0}380$ with a largest $K_0$ value of $380$~MeV. The spread in maximum central density among the interactions reflects the combined effects of the stiffness of the nuclear matter EOS and the momentum dependence of the mean field potential. The stiffer EOS (L$45$D$03$K$_{0}380$) produces a lower peak density due to stronger repulsion, whereas the momentum-independent potential (L$45$MID) leads to a higher compression. Such high-density conditions with approximately $2.5\rho_0$ using various interactions offer a unique opportunity to explore the high-density behaviors of nuclear matter EOS.

\subsection{Effects of the momentum-dependent nuclear mean-field potentials}
The momentum-dependent component of the single-particle potential constitutes a critical aspect of nuclear interactions.
It plays a pivotal role in the dynamics of heavy-ion collisions.
The momentum dependence remains an open question, even at the nuclear saturation density, especially for its isospin-dependent part, i.e., the symmetry potential $U_{\rm sym}$. Here, we aim to investigate the influence of the momentum dependence in both $U_{0}$ and $U_{\text{sym}}$ on the proton anisotropic flows employing the above constructed interactions L$45$D$03$FL, L$45$Dm$03$, and L$45$MID, with L$45$D$03$ as baseline.

\begin{center}
\begin{figure*}  
	\includegraphics[width=0.75\linewidth]{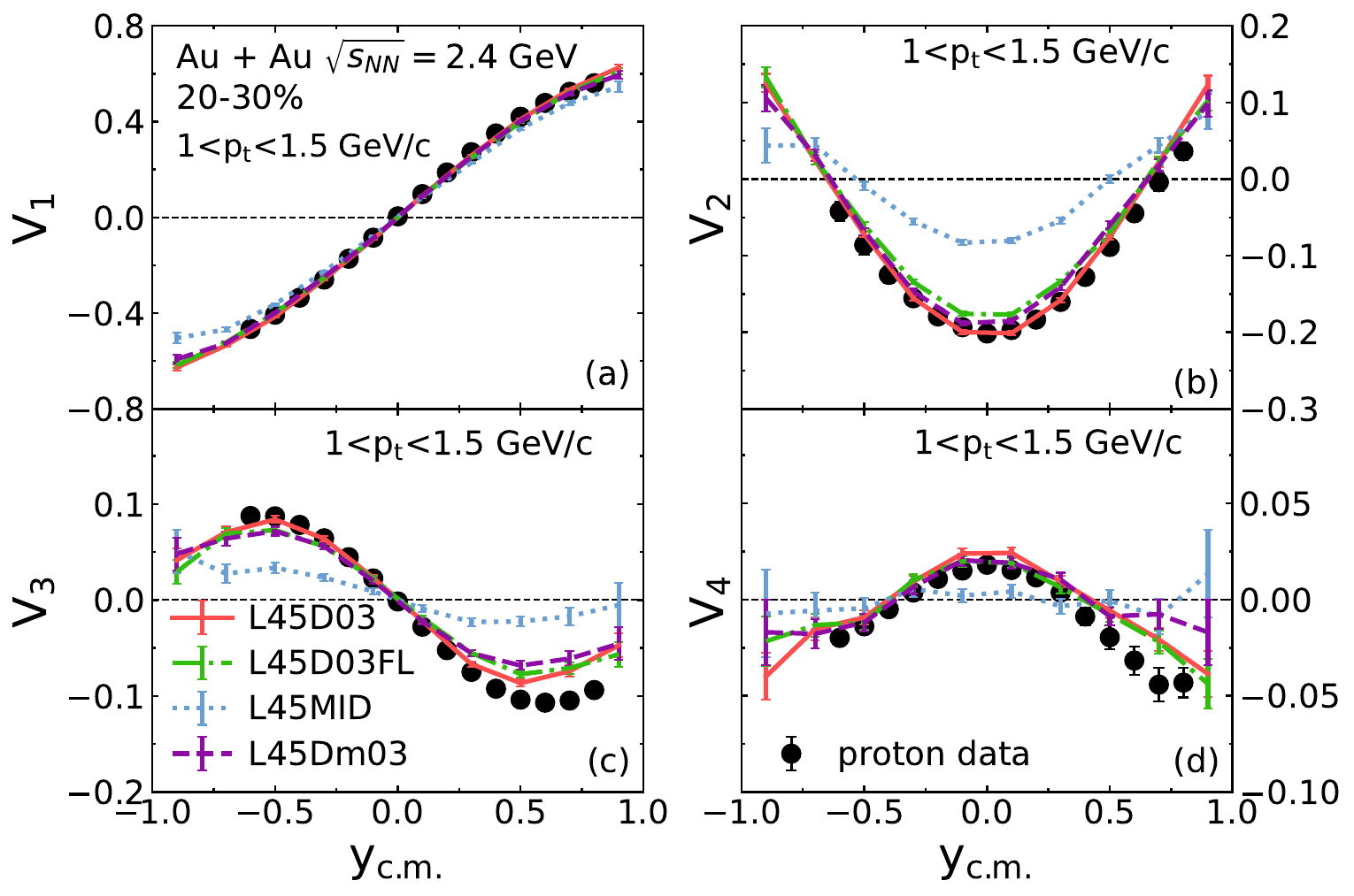}
	\caption{Directed ($v_1$) (a), elliptic ($v_2$) (b), triangular ($v_3$) (c) and quadrangular ($v_4$) (d) flows as functions of the center-of-mass rapidity $y_{\mathrm{c.m.}}$ for free protons in Au+Au collisions at $\sqrt{s_{\rm NN}}=2.4 \,\mathrm{GeV}$ (corresponding to $E_\mathrm{beam}$ $=$ $1.23A~\rm GeV$) predicted by the lattice BUU transport model with interactions L$45$D$03$, L$45$D$03$FL, L$45$Dm$03$ and L$45$MID. The corresponding HADES data~\cite{HADES:2020lob,HADES:2022osk} are also included for comparison.}
	\label{fig:U0MDrapall}
\end{figure*}
\end{center}

\begin{center}
\begin{figure*}
	\includegraphics[width=0.75\linewidth]{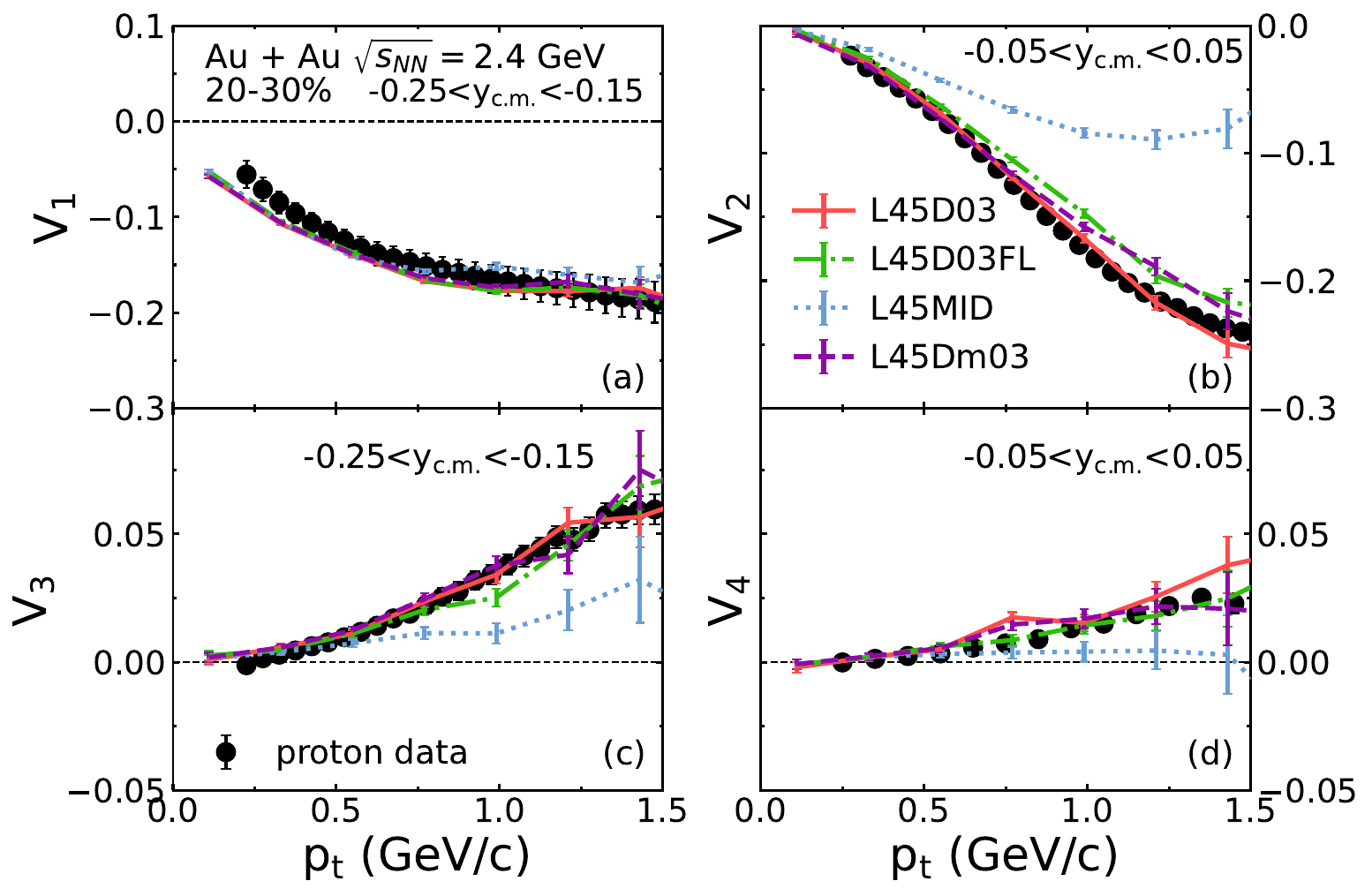}
	\caption{Directed ($v_1$) (a), elliptic ($v_2$) (b), triangular ($v_3$) (c) and quadrangular ($v_4$) (d) flows as functions of transverse momentum ($p_{\mathrm{t}}$) for protons in Au+Au collisions at $\sqrt{s_{\rm NN}}=2.4 \,\mathrm{GeV}$ (corresponding to $E_\mathrm{beam}$ $=$ $1.23A~\rm GeV$ ) predicted by lattice BUU transport model with L$45$D$03$, L$45$D$03$FL, L$45$Dm$03$ and L$45$MID interactions. The corresponding HADES data~\cite{HADES:2020lob,HADES:2022osk} are also included for comparison.}
	\label{fig:U0MDptall}
\end{figure*}
\end{center}

In order to demonstrate the impact of $U_{0}$ and $U_{\text{sym}}$, we present in Figs.~\ref{fig:U0MDrapall} and~\ref{fig:U0MDptall}, respectively, the rapidity  dependency and transverse momentum dependency, of the proton anisotropic flows $v_1$, $v_2$, $v_3$ and $v_4$, obtained within the LBUU transport model employing the four interactions L$45$D$03$, L$45$D$03$FL, L$45$Dm$03$, and L$45$MID.
The most notable feature in Figs.~\ref{fig:U0MDrapall} and \ref{fig:U0MDptall} is that the momentum-independent interaction L$45$MID systematically predicts weaker proton anisotropic flows $v_1$, $v_2$, $v_3$ and $v_4$ in both rapidity and transverse momentum dependencies. This overall flow suppression reflects a significantly weaker anisotropic compression with a smaller squeeze-out pressure, compared to the results from the baseline interaction L$45$D$03$. Such a discrepancy underscores the importance of momentum dependence in the nuclear mean-field potential for accurately describing the flow observables.

Anisotropic flows in heavy-ion collisions at HADES energy primarily originate from the interaction between the participant zone and the spectator remnants in non-central collisions. The spatial anisotropy of the initial overlap region is converted into momentum-space anisotropy mainly through geometric effects. Directed flow ($v_{1}$) arises from the sideward deflection or dragging of matter by the spectators, while elliptic flow ($v_{2}$) is dominated by the squeeze-out of participants due to blocking by the receding spectators. The triangular flow ($v_{3}$) and quadrangular flow ($v_{4}$), are mainly generated by the system’s collective response to the initial triangular and quadrangular shape fluctuations of the overlap region, respectively. Pressure gradients developed during the compression stage also contribute to the collective expansion and modulate the magnitude of all anisotropic flows. Momentum-dependent interactions such as L$45$D$03$ and L$45$D$03$FL enhance the repulsion for nucleons with higher momenta, thereby strengthening both the geometric and pressure-driven contributions to the flows.
In contrast, a momentum-independent potential weakens this driving mechanism. This is clearly demonstrated by the L$45$MID calculations, which systematically yield smaller flow values than the HADES data.
It is interesting to see from Figs.~\ref{fig:U0MDrapall} and \ref{fig:U0MDptall} that the baseline interaction L$45$D$03$ can give a reasonable description of the HADES data on the proton anisotropic flows except for the low-$p_t$ behaviors of $v_1$ (See Fig.~\ref{fig:U0MDptall}~(a)).
We therefore conclude that incorporating momentum dependence into the single-nucleon potential is essential for reproducing the anisotropic flow data reported by the HADES collaboration. It is worth noting that similar effects of momentum-dependent nucleon potentials have also been observed in heavy-ion collisions at lower beam energies~\cite{Danielewicz:1999zn}.

Apart from the observed strong effect on the proton anisotropic flows of eliminating the momentum-dependence in the nucleon single-particle potentials with the interaction L$45$MID, the impact of introducing a different momentum-dependent potential, namely replacing the Hama potential in the baseline interaction L$45$D$03$ by the Feldmeier and Lindner one (i.e., L$45$D$03$FL), or employing a different $U_{\rm sym}$ (i.e., L$45$Dm$03$), is found to be modest as evidenced in Figs.~\ref{fig:U0MDrapall} and \ref{fig:U0MDptall}.
For the interaction L$45$D$03$FL, since its $U_0$ for high-momentum nucleons is lower than that of L$45$D$03$, which suggests a weaker repulsion of high-momentum nucleons, it thus predicts a relatively weaker magnitude in $v_1$, $v_2$, $v_3$ and $v_4$, especially for $v_2$ at mid-rapidity and high transverse momentum region.
Similar effects are also observed in the case of the LBUU calculation using L$45$Dm$03$.
As illustrated in panel (b) of Fig.~\ref{fig:U0Usy}, the rising symmetry potential in L$45$Dm$03$ generates a weaker repulsive potential for protons compared to that of L$45$D$03$, especially at high momentum, thus leading to weaker proton anisotropic flows in L$45$Dm$03$. This feature indicates that the $p_t$-dependence of the proton $v_2$ exhibits a modest sensitivity to the momentum dependence of the symmetry potential.

Overall, the present LBUU transport model results with interactions L$45$D$03$, L$45$D$03$FL, L$45$Dm$03$ and L$45$MID clearly demonstrate that a less repulsive momentum-dependent proton potential will lead to weaker magnitude of the proton anisotropic flows in Au+Au collisions at HADES energy. These results highlight the importance of the momentum dependence in both $U_{0}$ and $U_{\text{sym}}$ for quantitatively describe the proton anisotropic flows in heavy-ion collisions at HADES energies.

\begin{center}
\begin{figure*}
	\includegraphics[width=0.75\linewidth]{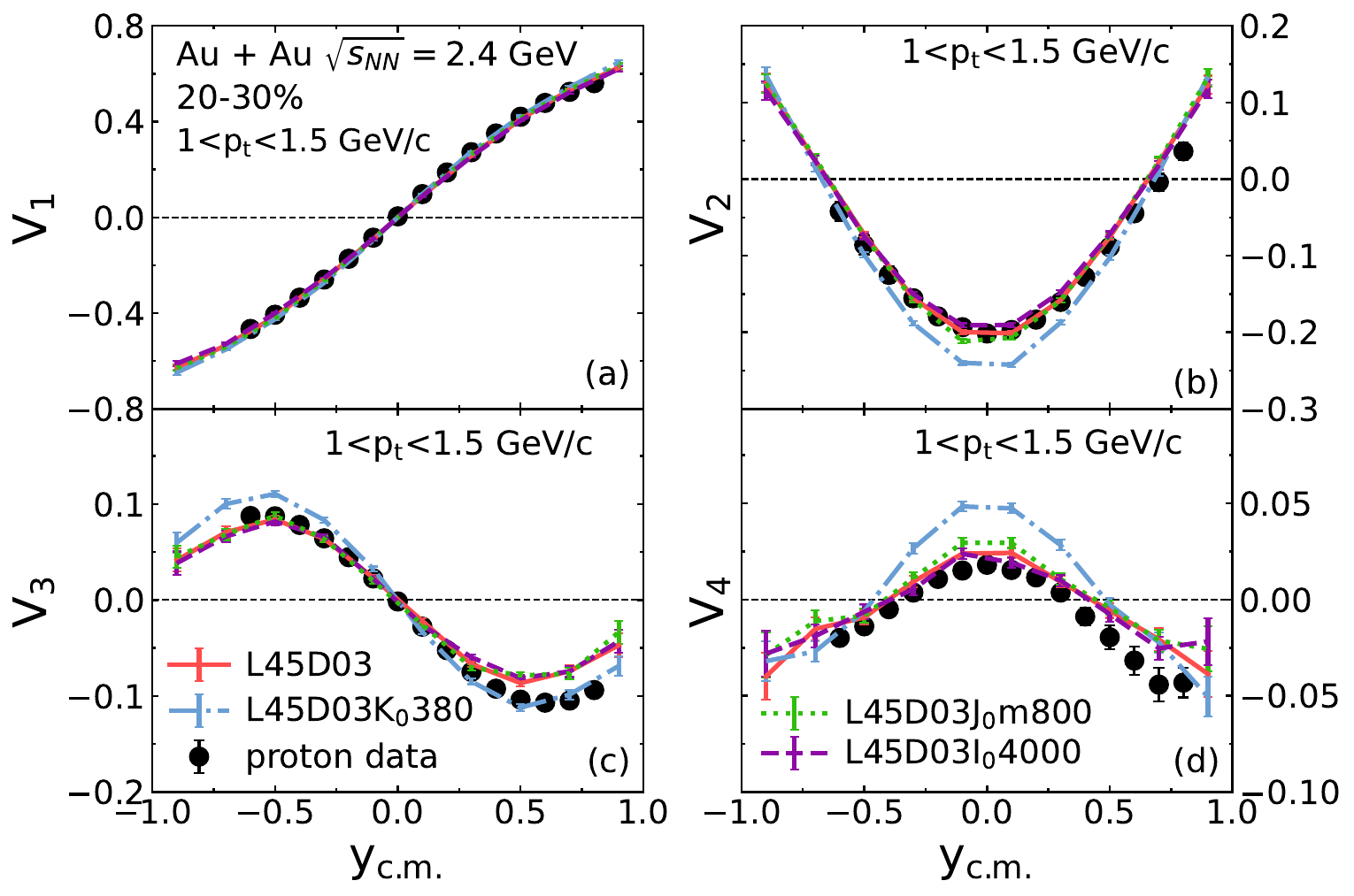}
	\caption{Same as Fig.~\ref{fig:U0MDrapall} but for interactions L$45$D$03$, L$45$D$03$K$_{0}380$, L$45$D$03$J$_{0}$m$800$ and L$45$D$03$I$_{0}4000$.}
	\label{fig:K0J0rapall}
\end{figure*}
\end{center}

\begin{center}
\begin{figure*}
	\includegraphics[width=0.75\linewidth]{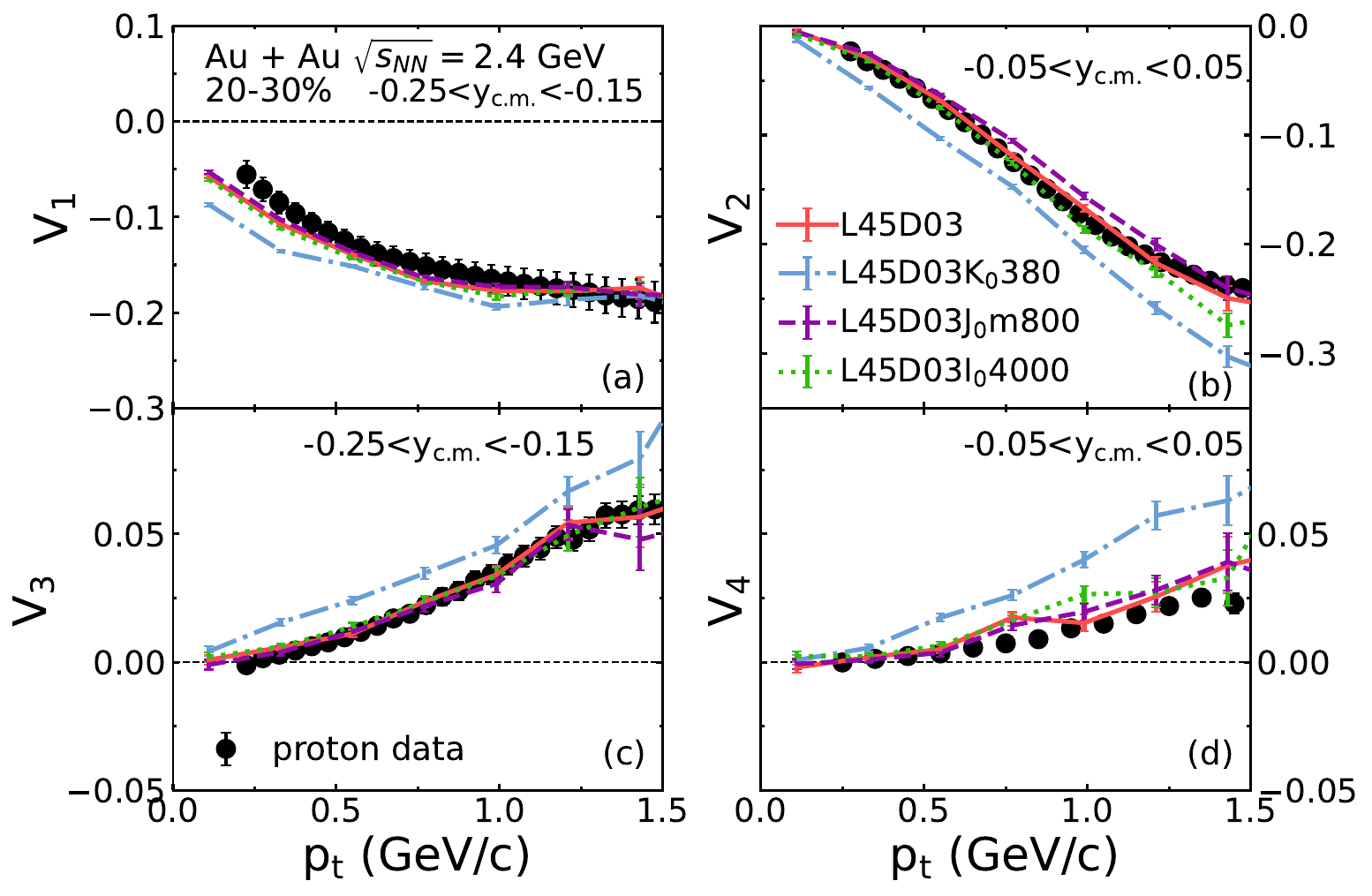}
	\caption{Same as Fig.~\ref{fig:U0MDptall} but for interactions L$45$D$03$, L$45$D$03$K$_{0}380$, L$45$D$03$J$_{0}$m$800$ and L$45$D$03$I$_{0}4000$.}
	\label{fig:K0J0ptall}
\end{figure*}
\end{center}

\subsection{Effects of the stiffness of symmetric nuclear matter EOS}
The stiffness of symmetric nuclear matter EOS plays a crucial role in understanding the dynamics of heavy-ion collisions, as it influences the pressure and density evolution of the matter produced in the collisions.
Around nuclear saturation density, the stiffness is typically characterized by the incompressibility coefficient $K_0$.
At supra-saturation densities, the influence of higher-order characteristic quantities---such as the skewness coefficient $J_0$ and the kurtosis coefficient $I_0$---could become increasingly significant.

To see the effects of $K_0$, $J_0$ and  $I_0$ on the proton anisotropic flows, we show in Figs.~\ref{fig:K0J0rapall} and~\ref{fig:K0J0ptall}, respectively, the rapidity and transverse momentum dependence of $v_1$, $v_2$, $v_3$, and $v_4$ of free protons in Au+Au collisions at $\sqrt{s_{\rm NN}}=2.4~\mathrm{GeV}$, as predicted by the LBUU transport model employing the interactions with different incompressibility coefficient $K_0$, skewness coefficient $J_0$ or kurtosis coefficient $I_{0}$, namely, L$45$D$03$, L$45$D$03$K$_{0}380$, L$45$D$03$J$_{0}$m$800$ and L$45$D$03$I$_{0}4000$.
The results show that the stiffness of the EOS of symmetric nuclear matter have strong impacts on the proton anisotropic flows for both rapidity and transverse momentum dependence. In particular, while the baseline interaction L$45$D$03$ (with $K_{0}$=$230$ MeV) provides a nice description of the rapidity and transverse momentum dependence across all flow coefficients ($v_1$ to $v_4$), the very stiff interaction L$45$D$03$K$_{0}380$ (with $K_{0}$=$380$ MeV) predicts a much stronger magnitude in $v_1$, $v_2$, $v_3$ and $v_4$, especially for $v_2$, $v_3$ and $v_4$ at mid-rapidity and high transverse momentum region compared with those from the L$45$D$03$. This enhancement can be understood from panel (e) of Fig.~\ref{fig:U0Usy}, from which one can see that L$45$D$03$K$_{0}380$ exhibits much larger pressure gradient than the baseline interaction L$45$D$03$, leading to a much stronger collective expansion and thus significantly larger magnitude of proton anisotropic flows.

On the other hand, it is seen from Figs.~\ref{fig:K0J0rapall} and~\ref{fig:K0J0ptall} that the higher-order EOS characteristic quantities $J_0$ and $I_0$ in interactions L$45$D$03$J${0}$m$800$ and L$45$D$03$I$_{0}4000$, respectively, exert a relatively modest influence on the proton anisotropic flows.
In particular, one can still see clear effects on $v_2$ at mid-rapidity and high transverse momentum.
The limited sensitivity of the proton anisotropic flows to $J_0$ and $I_0$ can be attributed to the fact that the higher-order parameters $J_0$ and $I_0$ primarily modify the behaviors of symmetric nuclear matter EOS at densities significantly higher than saturation density. Furthermore, the finite duration of such high-density phases in nuclear matter in the collision limit the integrated impacts of $J_0$ and $I_0$ on the global pressure gradient that drives the proton anisotropic flow, making their influence less pronounced than that of the dominant incompressibility coefficient $K_0$.
In addition, we note that the effects of high-order characteristic quantities might depend on their variation magnitudes relative to the baseline interaction. We emphasize that our present analysis is qualitative, and therefore, a more quantitative investigation of these higher-order characteristic quantities is important in the future research.

The above results demonstrate that the stiffness of the symmetric part of the nuclear matter EOS plays a crucial role in describing collective flow observables.
By systematically comparing the proton flow coefficients obtained with interactions characterized by different stiffness values of the symmetric nuclear matter EOS, insights into the properties of dense nuclear matter can be obtained. Specifically, it is interesting to see that an incompressibility value of $K_0 = 230$~MeV consistent with empirical value constrained by giant monopole resonance measurements in finite nuclei~\cite{Blaizot:1980tw,Youngblood:1999zza,Li:2007bp,Garg:2018uam,Li:2022suc} yields the best overall agreement with experimental prton flow data in Au+Au collisions at HADES.
Furthermore, our analysis highlights the non-negligible influence of higher-order density characteristic quantities in symmeric nuclear matter EOS for accurately describing the proton collective flow observables.
Indeed, while the skewness coefficient $J_{0}$ and the kurtosis coefficient $I_{0}$ have a relatively modest impact compared to $K_0$, the observed variations in the proton anisotropic flows emphasize the need of considering the higher-order characteristic quantities, when accurately analyzing collective flow data to probe the properties of nuclear matter, especially at high densities.

\begin{center}
\begin{figure*}
	\includegraphics[width=0.75\linewidth]{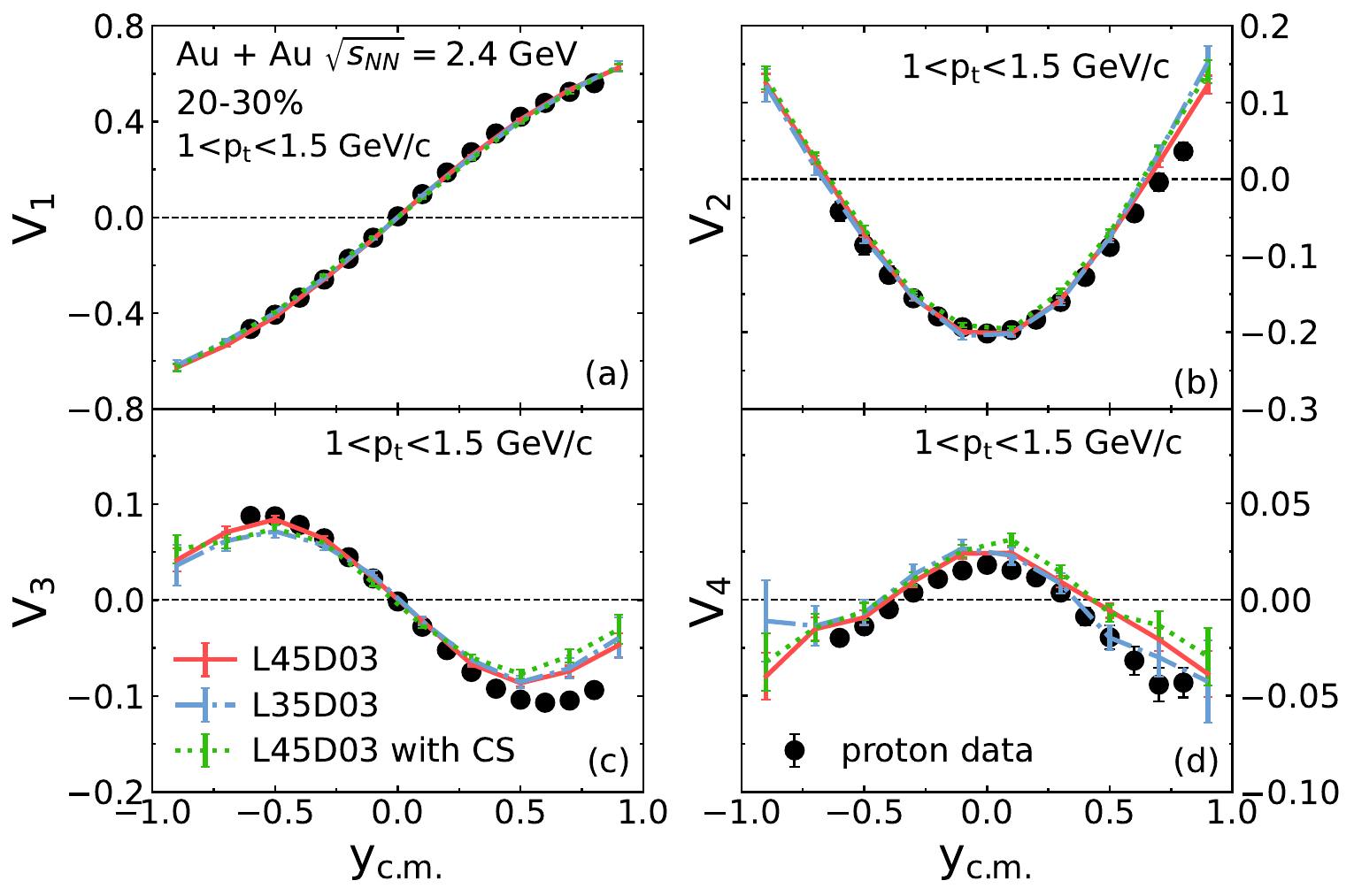}
	\caption{Same as Fig.~\ref{fig:U0MDrapall} but for interactions L$45$D$03$ (with and without in-medium modifications to nucleon-nucleon elastic cross sections) and L$35$D$03$.}
	\label{fig:CSrapall}
\end{figure*}
\end{center}

\begin{center}
\begin{figure*}
	\includegraphics[width=0.75\linewidth]{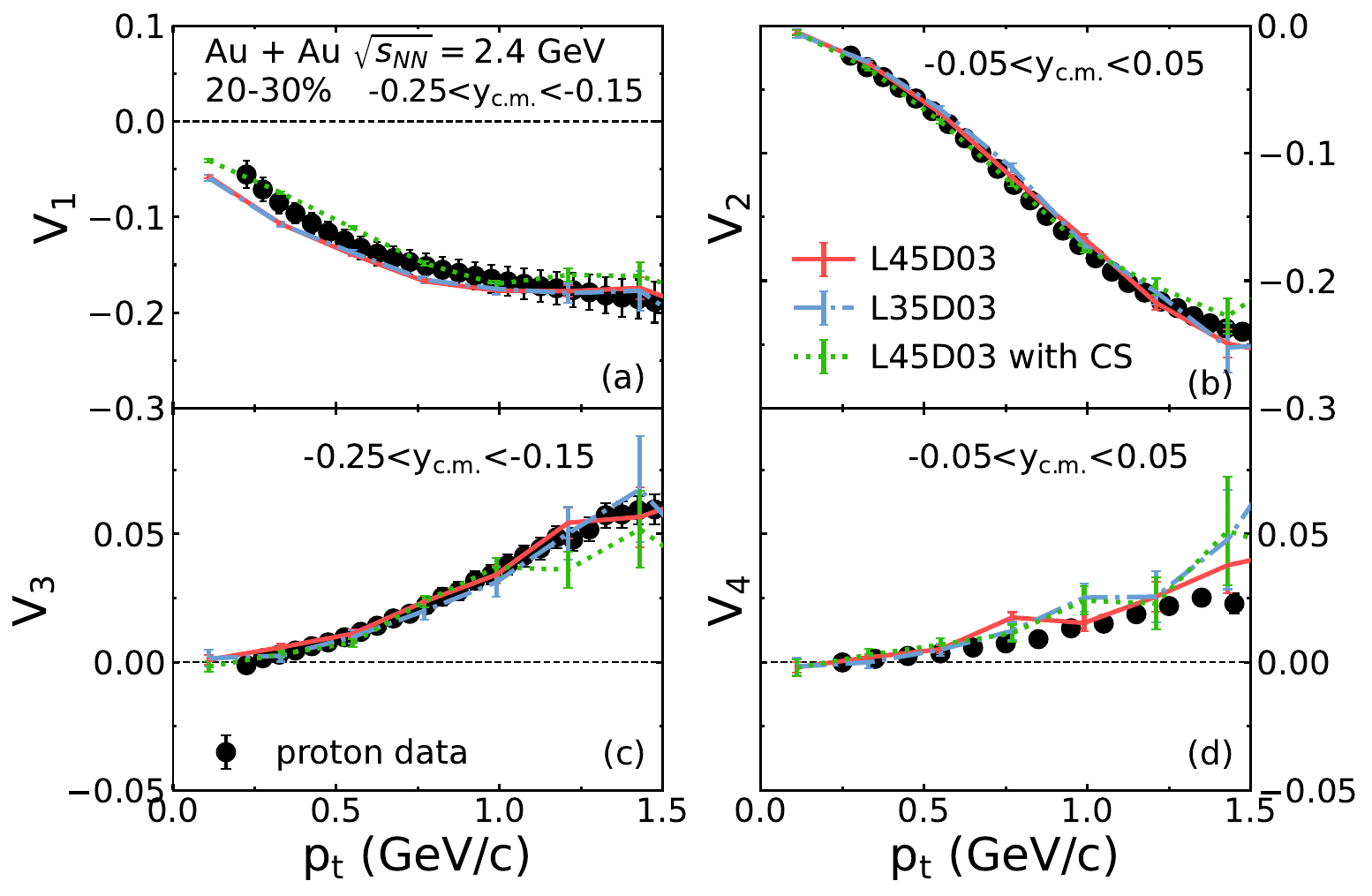}
	\caption{Same as Fig.~\ref{fig:U0MDptall} but for interactions L$45$D$03$ (with and without in-medium modifications to nucleon-nucleon elastic cross sections) and L$35$D$03$.}
	\label{fig:CSptall}
\end{figure*}
\end{center}

\subsection{Effects of the symmetry energy and in-medium nucleon-nucleon elastic cross sections}
\label{S:cs}
The symmetry energy, while generally considered to have a relatively smaller effect than the symmetric part of the nuclear matter EOS in heavy-ion collisions, exhibits significant uncertainties in the high-density regime.
These uncertainties stem from both theoretical calculation/modeling challenges and the limit of high-quality data of sensitive experimental probes.
Despite the relatively modest influence of the symmetry energy compared to the symmetric nuclear matter EOS, the precise characterization of the density dependence of the symmetry energy remains crucial for understanding the dense nuclear matter properties, particularly in the context of compact stars and heavy-in collisions induced by radioactive nuclei where isospin can span a wide range.
Here, we investigate the effects of the symmetry energy through the interaction L$35$D$03$, which has a softer density dependence of the symmetry energy compared with that of the baseline interactions L$45$D$03$ in the density range of around $1.5\rho_{0}$-$3\rho_{0}$ (see panel (f) of Fig.~\ref{fig:U0Usy}).
Shown in Figs. \ref{fig:CSrapall} and \ref{fig:CSptall} are, respectively, the rapidity and transverse momentum dependencies of the proton flow coefficients $v_1$, $v_2$, $v_3$ and $v_4$, as in Figs.~\ref{fig:U0MDrapall} and \ref{fig:U0MDptall} but with the L$45$D$03$ and L$35$D$03$ interactions.
It can be seen that the proton anisotropic flows overall exhibit limited sensitivity to the suprasaturation density behaviors of the symmetry energy, although some effects are visible for the higher-order $v_3$ and $v_4$.

Finally, we investigate the impact of the in-medium modification of the nucleon-nucleon elastic cross sections on the proton anisotropic flows.
The in-medium modifications can influence the nuclear stopping and nucleon collective flows, particularly at lower incident energies. As a result, these modifications will interplay with the nuclear matter EOS and nucleon effective masses in the reaction dynamics, thereby complicating the extraction of information of nuclear matter EOS or nucleon effective masses from heavy-ion collisions.
As shown in Fig.~\ref{fig:CS}, one can see that the in-medium correction leads to a significant suppression of the nucleon–nucleon elastic cross sections at HADES energy.

Also included in Figs.~\ref{fig:CSrapall} and \ref{fig:CSptall} are the corresponding results with the in-medium nucleon-nucleon elastic cross sections based on a parametrization form obtained from Eq.~(\ref{CS-form})~\cite{Wang:2020xgk} (i.e., L$45$D$03$ with CS).
One can see that although the parametrization Eq.~(\ref{CS-form}) leads to a significant in-medium correction on the nucleon-nucleon elastic cross sections, non-negligible changes can only be observed in the transverse momentum dependence of the directed flow $v_1$ as shown in Fig.~\ref{fig:CSptall}~(a).
The reduced magnitude of the directed flow can be understood as follows: larger cross sections without in-medium corrections result in more nucleon–nucleon scatterings, which deflect more nucleons into the transverse plane and consequently lead to stronger directed flow and enhanced nuclear stopping.
It is worth noting that there exists a close correlation between nuclear stopping and directed flow~\cite{FOPI:2004orn}.
On the other hand, the results for L$45$D$03$ with and without the in-medium cross section corrections yield nearly identical high-order flow coefficients, namely, $v_2$, $v_3$ and $v_4$.
Together with the observation in Figs.~\ref{fig:U0MDrapall}--\ref{fig:K0J0ptall}, namely that $v_2$, $v_3$ and $v_4$ are sensitive to both the single-nucleon potential and the stiffness of symmetric nuclear matter EOS, our present results suggest that the higher-order anisotropic flows could serve as effective probes of the nuclear matter EOS and the nucleon single-particle potential with less disturbation from the uncertainties of the in-medium nucleon-nucleon cross section correction.

The above results indicate that the high-density symmetry energy has relatively weak effects on proton anisotropic flow data in heavy-ion collisions at HADES energies. On the other hand, while the in-medium nucleon-nucleon elastic cross section modification exhibits a less pronounced overall impact, this modification plays a crucial role in precisely characterizing the flow data, particularly for the directed flow $v_{1}$. These results emphasize the essential requirement of incorporating the in-medium nucleon-nucleon cross section correction effects into theoretical frameworks to obtain a comprehensive and accurate description of the proton collective flow observables in heavy-ion collisions at HADES energy.
Our results also imply that
the higher-order anisotropic flows $v_2$, $v_3$ and $v_4$ may serve as good probes of the symmetric nuclear matter EOS and the nucleon single-particle potential, since the largely uncertain in-medium nucleon-nucleon cross section correction and high-density symmetry energy have relatively weak effects on these higher-order anisotropic flows.

\section{Summary}
\label{summary}

In this work, we have systematically investigated the impacts of several factors on proton anisotropic flows in Au+Au collisions at $\sqrt{s_{\rm NN}}$ $=$ $2.4~\rm{GeV}$, using a combined framework of the LBUU transport model and the N$5$LO Skyrme pseudopotential.
These factors include the momentum dependence of nucleon single-particle potential including the symmetry potential, the stiffness of symmetric nuclear matter EOS, the high-density behaviors of the symmetry energy, and the in-medium modifications on nucleon-nucleon elastic cross sections.
The flexibility in the density, momentum and isospin dependence of the N$5$LO Skyrme pseudopotential allows for independent examination of each of the above factors, while the advanced treatment used in solving the BUU equation in the LBUU transport model ensures very accurate descriptions of the collision dynamics.

Our findings reveal that the proton anisotropic flow coefficients exhibit significant sensitivity to the momentum dependence of nucleon single-particle potential and the incompressibility coefficient $K_0$ of the symmetric nuclear matter. In particular, it is found that incorporating momentum dependence into the nuclear single-particle potential is essential for reproducing the HADES flow data.
On the other hand, the higher-order coefficients of symmetric nuclear matter, namely the skewness coefficient $J_{0}$ and the kurtosis coefficient $I_{0}$ mainly exhibit a modest sensitivity to the transverse momentum dependence of the proton $v_2$.
Meanwhile, the high-density behaviors of the symmetry energy has limited effects on the proton anisotropic flows.
While the overall influence is relatively small, the in-medium modification of the nucleon-nucleon elastic cross sections nonetheless induces a noticeable effect on the transverse-momentum dependence of the proton $v_1$.

Given that the in-medium modifications of nucleon-nucleon cross sections and the high-density symmetry energy have a practically negligible influence on $v_2$, $v_3$ and $v_4$, these high-order flow coefficients may thus serve as good probes of the symmetric nuclear matter EOS and the momentum dependence of nucleon mean-field potential including the symmetry potential (and the associated isospin splitting of nucleon effective masses), largely free from the uncertainties introduced by the in-medium modifications of nucleon-nucleon cross sections and the high-density behaviors of the symmetry energy.

It should be emphasized that to globally describe the proton anisotropic flows $v_1$, $v_2$, $v_3$, and $v_4$, while the momentum dependence of nucleon single-particle potential and $K_0$ play predominant roles, the influences from higher-order EOS coefficients ($J_0$, $I_0$), the momentum dependence of the symmetry potential, and in-medium modifications of nucleon-nucleon cross sections remain non-negligible and need careful consideration. Consequently, a comprehensive analysis should take all these factors into account. This multi-factor dependence of the flow observables underscores the necessity for Bayesian-type analyses in the future studies, as these analyses provide a robust statistical framework for disentangling the intricate interplay among nuclear matter properties, single-particle dynamics, and their collective manifestations in heavy-ion collisions.


\begin{acknowledgments}
This work was supported in part by the National Natural Science Foundation of China under Grant Nos. 12235010, 12575139 and 12147101, the National SKA Program of China (Grant No. 2020SKA0120300), the Science and Technology Commission of Shanghai Municipality (Grant No. 23JC1402700), and the Natural Science Foundation of Henan Province (Grant No. 242300421048).
\end{acknowledgments}




\bibliography{HADESflow.bib}

\end{document}